\documentclass[twocolumn]{aastex701}
\usepackage{amsmath}
\usepackage{lipsum}
\usepackage{booktabs}
\usepackage{tabularx}
\usepackage{makecell}
\usepackage{caption}
\usepackage{threeparttable}
\usepackage{graphicx}
\usepackage[noabbrev]{cleveref}
\usepackage{ulem}

\submitjournal{ApJ}

\begin{document}

\title{Widespread Extended [\ion{C}{2}] Emission in High-Redshift Galaxies: Insights from the FIRE-2 Cosmological Zoom-in Simulations}

\author[0009-0004-1270-2373]{Lun-Jun Liu}
\affiliation{California Institute of Technology, 1200 E California Blvd., Pasadena, CA 91125, USA}
\email[show]{lliu@caltech.edu}
\correspondingauthor{Lun-Jun Liu}

\author[0000-0003-4070-497X]{Guochao Sun}
\altaffiliation{CIERA Fellow}
\affiliation{CIERA and Department of Physics and Astronomy, Northwestern University, 1800 Sherman Ave., Evanston, IL 60201, USA}
\email{}

\author[0000-0002-9382-9832]{Andreas~L. Faisst}
\affiliation{IPAC, California Institute of Technology, 1200 E. California Blvd., Pasadena, CA 91125, USA}
\email{}

\author[0000-0002-4900-6628]{Claude-André Faucher-Giguère}
\affiliation{CIERA and Department of Physics and Astronomy, Northwestern University, 1800 Sherman Ave., Evanston, IL 60201, USA}
\email{}

\author[0000-0002-3950-9598]{Adam Lidz}
\affiliation{Department of Physics and Astronomy, University of Pennsylvania, 209 S. 33rd St., Philadelphia, PA 19104, USA}
\email{}

\begin{abstract}

Recent ALMA observations reveal diffuse [\ion{C}{2}] emission (``[\ion{C}{2}] halos") extending to $\sim 10\,$kpc in galaxies at $4 < z < 6$. These measurements provide new insights into high-redshift galactic ecosystems and processes that drive metal enrichment on inner circumgalactic scales. To better understand the nature of [\ion{C}{2}] halos, we analyze a suite of high-redshift FIRE-2 simulations at $5 \leq z \leq 6$ in the stellar mass range of $10^{9}$--$10^{10.5}\,M_{\odot}$. By post-processing these simulations with three-dimensional dust radiative transfer and photoionization modeling, we generate synthetic images of [\ion{C}{2}] and UV continuum emission, from which we extract one-dimensional surface brightness profiles. Our results reproduce both the galaxy-integrated and spatial distribution of [\ion{C}{2}] and UV emission, capturing in particular the more extended profile of [\ion{C}{2}] emission. Comparing the time evolution of [\ion{C}{2}] halos with bursty star formation histories of the simulated galaxies, we find that [\ion{C}{2}] emission becomes more spatially extended following the decline of star formation rate in recent starburst episodes. This implies a strong correlation between extended [\ion{C}{2}] emission and bursty star formation, consistent with a key role for star formation-driven outflows in producing [\ion{C}{2}] halos---though the kinematics of [\ion{C}{2}]-emitting gas suggest that inflows and turbulent motions are also significant contributors. We also find a modest contribution from satellite galaxies to extended [\ion{C}{2}] emission. Our framework can be readily applied to predict the observability of [\ion{C}{2}] halos at higher redshifts and extended to create spatially resolved synthetic observations of other important emission lines, such as [\ion{O}{3}] and H$\alpha$.

\end{abstract}

\keywords{\uat{Hydrodynamical simulations}{767} --- \uat{High-redshift galaxies}{734} --- \uat{Galaxy evolution}{594} --- \uat{Galaxy formation}{595} --- \uat{Interstellar medium}{847} --- \uat{Circumgalactic medium}{1879}}

\section{Introduction} \label{sec:intro} 

Studying the morphology of galaxies in the early Universe provides crucial insights into the physics of galaxy formation. Rest-frame ultraviolet (UV)/optical observations by the James Webb Space Telescope (JWST) have determined the galaxy size and mass relation of early galaxies up to $z\sim9$ (e.g., \citealt{Ormerod24,Allen25,Miller25}), showing that star-forming galaxies generally become more compact toward higher redshifts, with effective radii $r_\mathrm{e} \lesssim 3\,$kpc at $z\sim5$ and $r_\mathrm{e} \lesssim 2\,$kpc at $z\sim7$. Whereas emission at rest-UV/optical wavelengths directly tracing star formation is often obscured by dust, far-infrared (FIR) emission is hardly susceptible to dust attenuation, thus providing a unique window into the morphology and size of early galaxies \citep{HodgedaCunha2020}. 

The $158\,{\rm \mu m}$ fine-structure emission line of singly ionized carbon (\ion{C}{2} or ${\rm C^{+}}$) is a promising probe of dust-obscured star-forming galaxies. As a dominant coolant of the neutral interstellar medium (ISM; see \citealt{HT99,Goldsmith12}), the [\ion{C}{2}] line is bright in typical star-forming galaxies (e.g., \citealt{Stacey91,Delooze14}), correlating with the cold molecular gas content and the star formation rate (SFR) over wide dynamic ranges \citep[e.g.,][]{Delooze14,HC15,Zanella18,DZ20,Schaerer20,Vizgan22}. It is a sensitive tracer of the thermal response of neutral gas in FUV-irradiated photodissociation regions (PDRs), where photoelectric heating from polycyclic aromatic hydrocarbon (PAH) and dust grains is balanced by FIR fine-structure cooling dominated by [\ion{C}{2}] \citep{Wolfire1995,HT99}. In some systems, shocks and turbulent dissipation can additionally power enhanced [\ion{C}{2}] emission \citep{Appleton2013,Lesaffre2013}, enabling [\ion{C}{2}] to probe both radiative and mechanical feedback linked to the formation of massive stars. Interferometric observatories such as the Atacama Large Millimeter/submillimeter Array (ALMA) have enabled detailed studies of high-redshift objects, particularly by constraining [\ion{C}{2}] up to $z\sim11$ (e.g., \citealt{Stacey10,Delooze14,HC15,Maiolino15,Matthee17,Smit18,Zanella18,Fujimoto19,LF20,Rizzo21,Bouwens22,Fudamoto24,Kaasinen24}).

Recent efforts have been made to investigate the spatial distribution of [\ion{C}{2}] emission in early star-forming galaxies. \cite{Fujimoto19} conducted a $uv$-visibility plane stacking analysis of 18 star-forming galaxies (${\rm SFR}\sim 10-70\,M_{\odot}/{\rm yr}$) at $z\sim 5-7$ from deep ALMA data and first reported the existence of extended [\ion{C}{2}] emission (namely, the [\ion{C}{2}] halo). This work demonstrated that the stacked radial profile of [\ion{C}{2}] surface brightness is significantly more extended than that of the UV continuum and reaches a distance of $\sim 10\,$kpc, far beyond the effective radius of the stacked galaxies ($r_{\rm e} = 1.1\,$kpc) and out to scales characteristic of their inner ($\sim 0.1-0.3R_\mathrm{vir}$) circumgalactic medium \citep[CGM; e.g.,][]{Tumlinson17,Stern21,Gurvich2023}.

The ALMA Large Program to Investigate C$^+$ at Early Times Survey \citep[ALPINE; \#2017.1.00428.L, PI: Le Fèvre;][]{LF20,Bethermin20,Faisst20} subsequently studied [\ion{C}{2}] halos in a large sample of $z\sim4-6$ galaxies. Leveraging this dataset, \cite{Ginolfi20a} identified $10\,$kpc-scale [\ion{C}{2}] emission with signatures of outflows through a stacking analysis of 50 [\ion{C}{2}]-detected galaxies, while \cite{Fujimoto20} found that the majority of isolated galaxies exhibit spatially extended [\ion{C}{2}] emission based on [\ion{C}{2}] size measurements of individual galaxies. Later ALMA observations of a subsample of massive main-sequence ALPINE galaxies \citep[CRISTAL; \#2021.1.00280.L, PI: Herrera-Camus;][]{HC25} further confirmed the extended nature of [\ion{C}{2}] emission at $z \sim 5$ \citep{Solimano24,Ikeda25,Birkin25,Posses25,Faisst2026a}. At higher redshifts, \cite{Fudamoto22} stacked 28 galaxies from the ALMA Reionization Era Bright Emission Line Survey \citep[REBELS; \#2019.1.01634.L, PI: Bouwens;][]{Bouwens22} at $z \sim 7$ and demonstrated that [\ion{C}{2}] emission is spatially more extended than the UV continuum. Additional detections of extended [\ion{C}{2}] emission were also reported by ALMA small programs \citep[e.g.,][]{HC21,Akins22,Meyer22,Lambert23}.

The ubiquity of [\ion{C}{2}] halos in early galaxies suggests efficient carbon enrichment on circumgalactic scales, offering new insight into high-$z$ galactic ecosystems. Understanding the physical origins, properties, and evolution patterns of these [\ion{C}{2}] halos is therefore a compelling area of investigation. Several possible scenarios have been proposed, including satellite galaxies, extended PDRs on the circumgalactic scale, circumgalactic HII regions, cold streams of inflowing gas, outflows induced by feedback mechanisms, and tidal stripping in merging systems \citep[e.g.,][]{Fujimoto19,Fujimoto20,Ginolfi20b}. The existence of these scenarios points to the potentially diverse ways [\ion{C}{2}] halos may reveal the physics of early galaxy formation.

Theoretical investigations are crucial for understanding the nature of [\ion{C}{2}] halos. Analytic and semi-analytic models provide key physical insights into the origin and nature of [\ion{C}{2}] halos. \cite{Pizzati20,Pizzati23} developed semi-analytic models focusing on starburst-driven cooling outflows and found that outflows are a favored interpretation of the observed [\ion{C}{2}] halos. Notably, outflows and inflows can be strongly regulated by a galaxy's star formation activity \citep{Muratov15,AA17}, which is found to be intense in recent observations of early star-forming galaxies \citep[e.g.,][]{Faisst19,Cole25,Looser25}. Recent cosmological simulations of [\ion{C}{2}] emission at high redshifts, however, have produced somewhat mixed results. Many studies underpredict the size of [\ion{C}{2}] halos \citep{Pallottini17,Arata19,Pallottini19,Pallottini22,Schimek24}, while others partially reproduce the observed extended [\ion{C}{2}] emission \citep{Munoz24,Khatri25}. For example, \cite{Pallottini17,Pallottini19,Pallottini22} and \cite{Arata19} systematically underpredicted [\ion{C}{2}] halo sizes relative to observations \citep[see][]{Fujimoto19,Wang26}. \cite{Schimek24} found only $\sim 10\%$ of [\ion{C}{2}] originates from the CGM, likely underpredicting the CGM contribution, while \cite{Munoz24} and \cite{Khatri25} attributed extended [\ion{C}{2}] emission primarily to satellites and/or outflows. Some of these simulations prioritize large cosmological volumes at the expense of the resolution needed to resolve the dense ISM and the complex physics that gives rise to [\ion{C}{2}] emission. Moreover, the temporal evolution of [\ion{C}{2}] halos in relation to stochastic star formation activity remains largely unexplored.

\begin{table*}[htbp!]
\centering
\caption{High-redshift FIRE-2 Simulations Used in This Work}
\label{tab:simtable}
\begin{threeparttable}
\begin{tabularx}{0.99\textwidth}{cccccccc}
\toprule
\toprule
Name 
& \hspace{2.6mm} $M_{\rm halo}$ {[}$M_{\odot}${]} 
& \hspace{2.6mm} $R_{\rm vir}$ {[}kpc{]} 
& \hspace{2.6mm} $M_{\star}$ {[}$M_{\odot}${]} 
& \hspace{2.6mm} ${\rm SFR}$ {[}$M_{\odot}/{\rm yr}${]} 
& \hspace{2.6mm} $M_{\rm UV,intrinsic}$ 
& \hspace{2.6mm} $M_{\rm UV,attenuated}$ 
& \hspace{2.6mm} $L_{\rm [CII]}$ {[}$L_{\odot}${]} \\
\midrule

{\tt z5m12b} 
& \hspace{2.6mm} $8.7 \times 10^{11}$ 
& \hspace{2.6mm} 51.2 
& \hspace{2.6mm} $2.6 \times 10^{10}$ 
& \hspace{2.6mm} 146.6 
& \hspace{2.6mm} $-24.1$ 
& \hspace{2.6mm} $-21.5$ 
& \hspace{2.6mm} $1.8 \times 10^{9}$ \\

{\tt z5m12c} 
& \hspace{2.6mm} $7.9 \times 10^{11}$ 
& \hspace{2.6mm} 49.5 
& \hspace{2.6mm} $1.8 \times 10^{10}$ 
& \hspace{2.6mm} 103.0 
& \hspace{2.6mm} $-23.4$ 
& \hspace{2.6mm} $-20.8$ 
& \hspace{2.6mm} $1.8 \times 10^{9}$ \\

{\tt z5m12d} 
& \hspace{2.6mm} $5.7 \times 10^{11}$ 
& \hspace{2.6mm} 44.5 
& \hspace{2.6mm} $1.2 \times 10^{10}$ 
& \hspace{2.6mm} 11.6 
& \hspace{2.6mm} $-21.8$ 
& \hspace{2.6mm} $-20.8$ 
& \hspace{2.6mm} $5.6 \times 10^{8}$ \\

{\tt z5m12e} 
& \hspace{2.6mm} $5.0 \times 10^{11}$ 
& \hspace{2.6mm} 42.6 
& \hspace{2.6mm} $1.4 \times 10^{10}$ 
& \hspace{2.6mm} 98.9 
& \hspace{2.6mm} $-23.7$ 
& \hspace{2.6mm} $-21.5$ 
& \hspace{2.6mm} $4.3 \times 10^{8}$ \\

{\tt z5m12a} 
& \hspace{2.6mm} $4.5 \times 10^{11}$ 
& \hspace{2.6mm} 41.1 
& \hspace{2.6mm} $5.4 \times 10^{9}$ 
& \hspace{2.6mm} 14.0 
& \hspace{2.6mm} $-21.6$ 
& \hspace{2.6mm} $-20.2$ 
& \hspace{2.6mm} $3.2 \times 10^{8}$ \\

{\tt z5m11f} 
& \hspace{2.6mm} $3.1 \times 10^{11}$ 
& \hspace{2.6mm} 36.4 
& \hspace{2.6mm} $4.7 \times 10^{9}$ 
& \hspace{2.6mm} 24.8 
& \hspace{2.6mm} $-23.1$ 
& \hspace{2.6mm} $-21.9$ 
& \hspace{2.6mm} $4.0 \times 10^{8}$ \\

{\tt z5m11e} 
& \hspace{2.6mm} $2.5 \times 10^{11}$ 
& \hspace{2.6mm} 33.6 
& \hspace{2.6mm} $2.5 \times 10^{9}$ 
& \hspace{2.6mm} 1.4 
& \hspace{2.6mm} $-19.4$ 
& \hspace{2.6mm} $-19.0$ 
& \hspace{2.6mm} $1.2 \times 10^{8}$ \\
\bottomrule
\end{tabularx}
\begin{tablenotes}[para,flushleft]
\footnotesize
\item \textbf{Notes.} Properties at $z = 5$ and post-processing results of simulated galaxies from the High-Redshift suite of FIRE-2 simulations \citep{Ma18,Ma19,Wetzel23,Wetzel25}. Only the primary galaxy/halo at the center of each zoom-in region is considered. Name: simulation ID. $M_{\rm halo}$: halo mass. $R_{\rm vir}$: virial radius. $M_{\star}$: stellar mass. SFR: averaged over the past $10\,$Myr. $M_{\rm UV,intrinsic}$ and $M_{\rm UV,attenuated}$: the intrinsic and dust-attenuated UV magnitudes. $L_{\rm [CII]}$: total [\ion{C}{2}] luminosity.
\end{tablenotes}
\end{threeparttable}
\end{table*}

The Feedback in Realistic Environments\footnote{\url{https://fire.northwestern.edu/}} \citep[FIRE;][]{Hopkins14,Hopkins18,Hopkins23} cosmological zoom-in simulations are uniquely positioned to address these challenges. FIRE provides a self-consistent cosmological framework while achieving sub-parsec spatial resolution (minimum adaptive softening $\epsilon_{\rm gas,min} \leq 0.42\,$pc), sufficient to resolve sub-structures within molecular clouds and capture the conditions and processes that produce [\ion{C}{2}] emission\footnote{We note that the [\ion{C}{2}]-emitting gas in the cold inner CGM is typically much better resolved than the hot diffuse CGM and thus should be distinguished from it.}. This resolution, combined with detailed prescriptions for stellar feedback, star formation, and chemical evolution, enables FIRE to model how feedback mechanisms shape gas distribution and kinematics on scales directly comparable to spatially resolved ALMA observations. Moreover, the high-cadence snapshots (every $\sim 15\,$Myr between $z = 5-6$) allow us to track the evolution of [\ion{C}{2}] emission in relation to time variability of galaxy-scale star formation.

In this work, we simulate [\ion{C}{2}] emission in high-resolution $5 \leq z \leq 6$ galaxies from the publicly released High-Redshift suite of FIRE-2 simulations \citep{Ma18,Ma19,Wetzel23,Wetzel25}. We implement a post-processing framework for simulated galaxies that builds on the method introduced by \cite{Liang24}. This framework enables morphological studies of simulated [\ion{C}{2}] emission and UV continuum, from which we classify extended vs. concentrated [\ion{C}{2}] emission and perform a direct comparison of radial surface brightness profiles from mock images with observations. To understand the nature of [\ion{C}{2}] halos, we investigate their evolution in the context of the UV continuum and star formation histories (SFH), through which we reveal the strong correlation between [\ion{C}{2}] halos and bursty star formation in high-$z$ galaxies. 

This paper is structured as follows. In Section~\ref{sec:methods}, we introduce the FIRE-2 simulation and describe our post-processing framework, including three-dimensional dust radiative transfer and photoionization modeling. In Section~\ref{sec:compare}, we present simulated galaxy properties and radial profiles of [\ion{C}{2}] and the UV continuum, which are in agreement with observational constraints. In Section~\ref{sec:explain}, we show the time evolution of simulated [\ion{C}{2}] halos to discuss their correlation with bursty SFHs, along with their physical drivers. We discuss the implications of our findings, followed by a future outlook in Section~\ref{sec:discussion}, and our conclusions are provided in Section~\ref{sec:conclusion}. Throughout, we assume cosmological parameters consistent with measurements by \cite{Planck16}. Magnitudes are given in the AB system \citep{Oke83}.

\section{Simulations and Methods} \label{sec:methods}

\subsection{The Simulations} \label{subsec:fire2}

In this work, we use a sample of seven simulated galaxies from the \emph{High Redshift} suite of the FIRE-2 simulations \citep{Ma18,Ma19,Wetzel23,Wetzel25}, including \texttt{z5m12a,b,c,d,e} and \texttt{z5m11e,f}. Their properties, including results from our post-processing analysis (discussed in Section~\ref{subsec:skirt} and Section~\ref{subsec:Cloudy}), at $z=5$ are shown in Table~\ref{tab:simtable}. These cosmological zoom-in simulations implement the \textsc{gizmo} code\footnote{\url{http://www.tapir.caltech.edu/~phopkins/Site/GIZMO.html}} in meshless finite-mass (MFM) mode with a typical baryonic particle mass of $m_{i} = 7100 \, M_{\odot}$ \citep{Hopkins15,Hopkins18,Wetzel23}. For each snapshot of the FIRE-2 simulations, dark matter (DM) halos are identified using \textsc{rockstar-galaxies} \footnote{\url{bitbucket.org/awetzel/rockstar-galaxies}} \citep{Behroozi13}.

The FIRE-2 physics model \citep{Hopkins18,Ma18,Ma19,Wetzel23,Wetzel25} incorporates radiative cooling over $10-10^{10}\,$K with metallicity-dependent fine structure and molecular cooling, and tracks 11 species (H, He, C, N, O, Ne, Mg, Si, S, Ca, Fe). Photoionization and heating from a redshift-dependent, spatially uniform UV background \citep{Faucher09}\footnote{The exact UVB model used is available at \url{https://galaxies.northwestern.edu/uvb-fg09}.} are included. Notably, the FIRE-2 simulations do not include black hole physics and active galactic nuclei (AGN) feedback in order to provide a clean, well-controlled baseline for studying the role of stellar feedback.

Star formation occurs in dense ($n_{\rm gas} > 1000 \, {\rm cm^{-3}}$), self-gravitating, self-shielding molecular gas at 100\% efficiency per local free-fall time. The model includes various realistic stellar feedback channels to capture the chemical evolution and processes of the ISM, including radiation pressure, stellar winds from OB and AGB stars, photoionization and photoelectric heating, and supernovae (Type II and Ia) injection of energy, momentum, mass, and metal. The source of these feedback mechanisms, each star particle in the FIRE-2 models is treated as a single stellar population with determined age and metallicity, with feedback rates and energies derived from \textsc{starburst99}\footnote{\url{https://massivestars.stsci.edu/starburst99/docs/default.htm}} stellar evolution models \citep{Leitherer99,Leitherer14} assuming a \cite{Kroupa01} initial mass function.

\begin{figure*}[ht!]
\centering
\includegraphics[clip, trim={0 0 0 4.1cm}, width=0.99\textwidth]{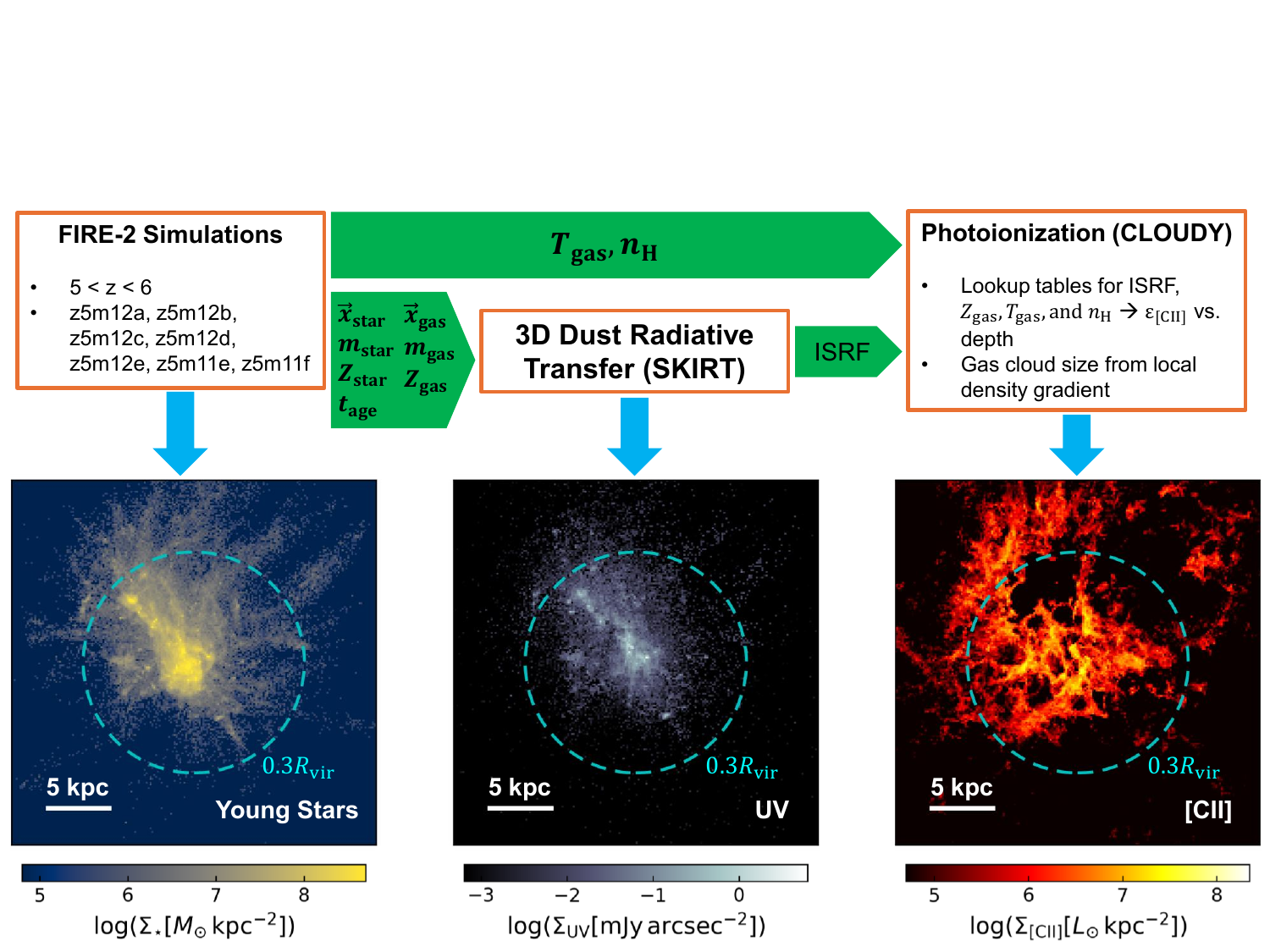}
\caption{Overview of our post-processing framework. The upper flowchart outlines key steps from left to right in orange blocks, and key physical information utilized at each step in green arrows. An example outcome of each step is shown for a FIRE-2 simulated galaxy ({\tt z5m12e}, with $M_{\star} = 10^{9.9}\,M_{\odot}$ at $z=5.77$). Left panel: Stellar mass surface density, $\Sigma_{\star}$, of young stars (age $<0.1$\,Gyr) extracted from the FIRE-2 simulations. Middle panel: Surface brightness of UV continuum, $\Sigma_{\mathrm{UV}}$, at $1500 \, \mathrm{\AA}$ modeled using the \textsc{skirt} three-dimensional dust radiative transfer. Right panel: Surface brightness of [\ion{C}{2}], $\Sigma_{\mathrm{[CII]}}$, obtained from photoionization modeling using \textsc{Cloudy}, which generates a lookup table of the [\ion{C}{2}] emissivity $\epsilon_{\mathrm{[CII]}}$ vs. depth into individual gas clouds covering a four-dimensional parameter space. The [\ion{C}{2}] luminosity, $L_{\mathrm{[CII]}}$, of individual gas clouds of a FIRE-2 galaxy can thus be obtained by integrating $\epsilon_{\mathrm{[CII]}}$ over the size of the gas cloud. The cyan dashed circles indicate the reference scale for our [\ion{C}{2}] halo analysis ($0.3R_{\rm vir}$, corresponding to $\sim 10\,$kpc for our galaxy sample).
\label{fig:2d_images}}
\end{figure*}

The FIRE models and simulations have succeeded in reproducing a broad range of observations at high redshifts ($z \geq 5$), such as UV luminosity functions and the UV-based cosmic star formation rate density \citep{Ma19,Sun23,Feldmann25}, the mass--metallicity relation \citep{Marszewski24,Marszewski2025}, and scaling relations of [\ion{C}{2}] with various galaxy properties \citep{Liang24}.

Our sample is selected to be broadly comparable to galaxies from the ALPINE and CRISTAL survey \citep{LF20,Bethermin20,Faisst20,Fujimoto20,HC25}, which characterize the statistics of [\ion{C}{2}] halos high-$z$ galaxies. Specifically, (i) the stellar mass ($M_{\star}$) and SFR of these selected FIRE-2 galaxies are broadly comparable to the observations, where $M_{\star} \sim 10^{9} - 10^{10.5} \, M_{\odot}$ and SFR$\, \sim 10 - 300 \,$$M_{\odot}$/yr, and (ii) snapshots in $5 \leq z \leq 6$ are selected in our investigation. Each simulated galaxy has 16 snapshots in $5 \leq z \leq 6$, and we compile a sample of 112 snapshots. In particular, we focus on gas within $0.1-0.3R_{\rm vir}$ as a proxy for the inner-CGM regime. This corresponds to the central $\sim 3-10\,{\rm kpc}$ region of these selected galaxies, comparable to the spatial extent of the observed [\ion{C}{2}] halos.

\subsection{Three-Dimensional Dust Radiative Transfer} \label{subsec:skirt}

An overview of the post-processing framework and example results are presented in Figure~\ref{fig:2d_images}. In this framework, we use the public three-dimensional (3D) Monte Carlo dust radiative transfer (RT) code \textsc{skirt}\footnote{\url{https://skirt.ugent.be/root/_home.html}} \citep{Baes15,Camps15,Camps20} to predict the UV continuum (rest-frame $\lambda = 1500\,$\AA) of simulated galaxies and the strength of the local interstellar interstellar radiation field (ISRF; in units of $G_0$\footnote{$1 \, G_{0} = 1\,\mathrm{Habing} = 1.6 \times10^{-3} \, {\rm erg \, s^{-1} \, cm^{-2}}$, which is the observed value in the solar neighborhood \citep{Habing68}.}) associated with individual gas particles, where the latter is a key parameter to predict the [\ion{C}{2}] luminosity ($L_{\rm [CII]}$) of these galaxies (Section~\ref{subsec:Cloudy}). \textsc{skirt} provides the full treatment of physical processes including scattering, absorption, and emission by dust, and can self-consistently calculate the ISRF distribution and spectral energy distributions (SEDs).

Our \textsc{skirt} post-processing framework builds upon the methods developed by \cite{Ma19} and \cite{Liang24}. We note that a recent work by \cite{Choban2025} explored more detailed dust modeling in the FIRE simulations through an evolution network on-the-fly, though we adopt a simpler approach here given our focus on [\ion{C}{2}] emission. Here we summarize key setups of our \textsc{skirt} implementation and refer interested readers to the aforementioned work for more details.

In this framework, we treat each star particle as a single stellar population defined by its initial mass, $Z_{\rm star}$, and $t_{\rm age}$ to generate its stellar emission spectrum from the \textsc{starburst99} SED libraries \citep{Leitherer99,Jonsson10}. The RT calculations are conducted on a 350-point wavelength grid equally spaced in logarithmic scale over $\lambda = 0.01$--$1000 \, \mu$m. At each wavelength grid point, $5\times10^{6}$ photon packets are launched for both the stellar emission phase and the subsequent dust re-emission phase. The simulation is iterated until numerical convergence is achieved. Synthetic images and SEDs of the simulated galaxies are then generated by placing virtual detectors at a distance of $10\,$Mpc, with the field of view extending out to $R_{\rm vir}$. These detectors can be oriented along multiple viewing angles to capture spatially resolved and integrated fluxes across the wavelength grid. We place three virtual detectors along orthogonal sightlines to mitigate the bias caused by using only a single sightline.

We take a simple approach by assuming that dust mass traces the metal mass within galaxies (e.g., \citealt{Camps16,Trayford17,Ma19,Liang21,Liang24}) and adopt a constant dust-to-metal mass ratio of $R_{\rm dust} = 0.4$ \citep{Dwek98} in gas with temperatures below $10^{6}\,$K, while considering gas at higher temperatures to be dust-free. The dust properties are modeled according to \cite{Weingartner01} with the Milky Way (MW) grain size distribution with an extinction curve parameter of $R_{\rm V} = 3.1$. The dust grid is reconstructed from gas particles using the built-in octree grid of \textsc{skirt} \citep{Saftly13,Saftly14}, with recursive subdivision of cells until each contains less than $3 \times 10^{-6}$ of the total dust mass and the optical depth of the $V$-band ($0.55\,\mu$m) becomes $< 1$. The self-absorption and re-emission of dust are self-consistently calculated by accounting for the non-local thermal equilibrium \citep{Baes11,Camps15}. In addition, we apply a temperature correction for dust in the \textsc{skirt} setting to account for dust heating by the cosmic microwave background \citep[CMB;][]{DC13}.

The \textsc{skirt} code outputs the UV continuum surface brightness ($\Sigma_{\rm UV}$) image of FIRE-2 galaxies and the specific interstellar radiation field (ISRF) $J_{\lambda}$ at sampled wavelengths for individual gas particles. An example $\Sigma_{\rm UV}$ image is illustrated in Figure~\ref{fig:2d_images} (middle panel), confirming that the UV continuum traces young stellar populations (age $< 0.1\,$Gyr). For each gas particle, we integrate $J_{\lambda}$ over the Habing band ($h \nu = 6$--$13.6\,$eV) and solid angle to obtain the far-UV ISRF strength ($J_{\rm FUV}$). This far-UV radiation singly ionizes carbon and photoelectrically heats the gas to excite ${\rm C^{+}}$ through collisions, resulting in the production of [\ion{C}{2}] emission (Section~\ref{subsec:Cloudy}) that balances gas heating.

\subsection{Modeling [\ion{C}{2}] with \textsc{Cloudy}} \label{subsec:Cloudy}

We predict the [\ion{C}{2}] luminosity of individual gas particles using the photoionization modeling code \textsc{Cloudy}\footnote{\url{https://gitlab.nublado.org/Cloudy/Cloudy}} version C17 \citep{Ferland17}. Given the gas properties and incident radiation field, \textsc{Cloudy} solves for the thermal, ionization, and chemical state of the gas and predicts the resulting emission and absorption spectra. This allows us to reliably estimate the [\ion{C}{2}] emissivity associated with the local physical conditions of each simulated gas particle.

Assuming the spherical symmetry of individual gas particles, the [\ion{C}{2}] luminosity of these idealized spherical `gas clouds' ($L_{\mathrm{[CII],cl}}$) can be computed based on its properties from the FIRE-2 galaxies, including gas mass of the cloud ($m_{\rm gas}$), gas-phase metallicity ($Z_{\rm gas}$), temperature of the gas cloud ($T_{\rm gas}$), and hydrogen nuclei number density ($n_{\rm H}$), and modeled $J_{\rm FUV}$ property from the \textsc{skirt} post-processing (as illustrated in Figure~\ref{fig:2d_images}). We adopt a fixed $T_{\rm gas}$ taken from the FIRE simulations for each gas cloud rather than allowing it to vary with depth into the gas cloud as in \textsc{Cloudy}, but we have verified that this causes only minor differences in $L_{\mathrm{[CII]}}$ \citep[see also][]{Lupi20}. Technically, \textsc{Cloudy} outputs the [\ion{C}{2}] volume emissivity ($\epsilon_{\mathrm{[CII]}} = \epsilon_{\mathrm{[CII]}}(r) \, [\mathrm{erg \, s^{-1} \, cm^{-3}}]$) at various radii of a gas cloud. Given that [\ion{C}{2}] is optically thin in typical ISM condition with density spans $1-10^{3}\,{\rm cm^{-3}}$ \citep[e.g.,][]{Goldsmith12,Olsen17}, $L_{\mathrm{[CII],cl}}$ can be calculated by integrating $\epsilon_{\mathrm{[CII]}}$ over the spherical volume of this gas cloud (with radius $R_{\rm cl}$), that is,

\begin{equation}
L_{\mathrm{[CII],cl}} = \int_{0}^{R_{\mathrm{cl}}} 4 \pi \epsilon_{\mathrm{[CII]}} r^{2} d r,
\end{equation}
For FIRE-2 zoomed galaxies, $R_{\rm cl}$ of each gas cloud can be approximated by the Sobolev-like length scale \citep{Sobolev57,Gnedin09,Gnedin11,Liang24} using the density gradient near this gas cloud:

\begin{equation}
R_{\mathrm{cl}} \sim \frac{\rho_{\mathrm{cl}}}{2 \, \langle \nabla \rho \rangle_{\mathrm{neighbors}}},
\end{equation}
where $\langle \nabla \rho \rangle_{\mathrm{neighbors}}$ is the average density gradient near a gas cloud with respect to its nearest 32 neighbors. The density gradient between a gas cloud and its $i$-th neighbor ($\nabla \rho_{\mathrm{cl},i}$) can be calculated using known properties of simulated galaxies: 

\begin{equation}
\nabla \rho_{\mathrm{cl},i} = \frac{\rho_{\mathrm{cl}} - \rho_{i}}{\Delta r_{\mathrm{cl},i}},
\end{equation}
where $\rho_{\mathrm{cl}}$ is the mass density of the gas particle, $\rho_{i}$ is the mass density of the $i$-th neighboring gas particle, and $\Delta r_{\mathrm{cl},i}$ is the distance in between.

Each gas cloud has its own unique set of parameters ($J_{\rm FUV}$, $Z_{\rm gas}$, $n_{\rm H}$, $T_{\rm gas}$); however, it is computationally unaffordable to run \textsc{Cloudy} case-by-case to obtain $L_{\mathrm{[CII],cl}}$ of each gas cloud. Instead, we generate a lookup table for a grid of these parameters and perform multivariate interpolation to interpolate $L_{\mathrm{[CII],cl}}$.

Specifically, we build our \textsc{Cloudy} model grid by varying the parameters as follows:

\begin{itemize}
    \item $\log(J_{\rm FUV}/G_{0})$ from $-1$ to $5$, in increments of $1$.
    \item $\log(Z_{\rm gas}/Z_{\odot})$ from $-2$ to $0.4$, in increments of $0.8$.
    \item $\log(n_{\rm H} \, {\rm cm^{-3}})$ from $-1$ to $5$, in increments of $1$.
    \item $\log(T_{\rm gas}/{\rm K})$ from $1$ to $5$, in increments of $0.5$.
\end{itemize}

\begin{figure}[h!]
\centering
\includegraphics[width=0.99\columnwidth]{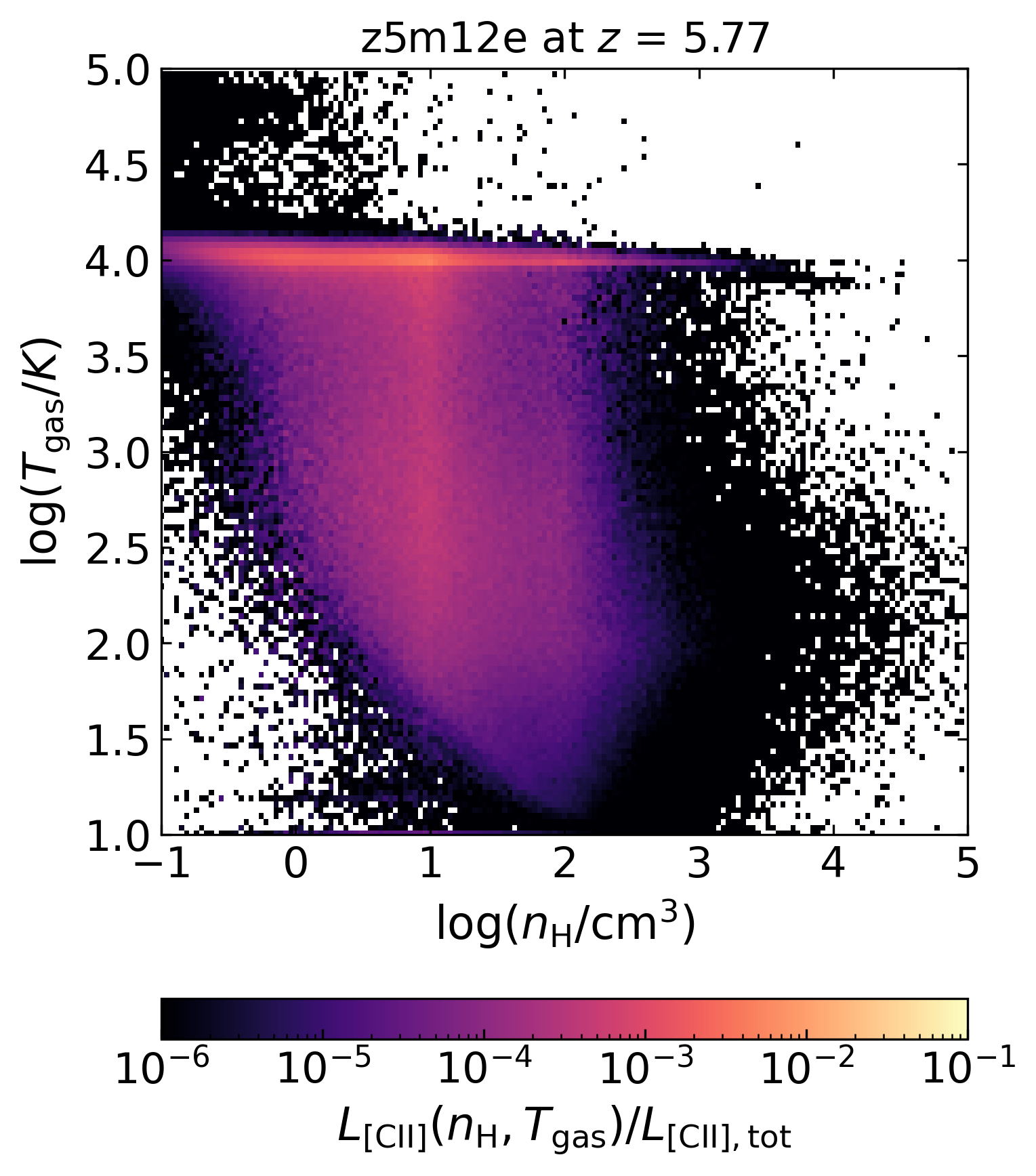}
\caption{Phase diagram of gas temperature ($T_{\rm gas}$) vs. density ($n_{\rm H}$) for an example FIRE-2 galaxy, {\tt z5m12e}, at $z = 5.77$ ($M_{\star} \sim 10^{9.9}\,M_{\odot}$), weighted and color-coded by $L_{\rm [CII]}(n_{\rm H}, T_{\rm gas})/L_{\rm [CII],tot}$, the fractional contribution to the total $L_{\rm [CII]}$ from gas clouds with certain $T_{\rm gas}$ and $n_{\rm H}$. The diagram includes all gas clouds within $0.3R_{\rm vir}$.
\label{fig:phase}}
\end{figure}

We select these grid boundaries to encompass the physical conditions under which most [\ion{C}{2}] emission is produced. Gas particles in our simulations with physical conditions outside this grid contribute negligibly to the total [\ion{C}{2}] luminosity (see Figure~\ref{fig:phase}), so we set their luminosities to zero.

In each \textsc{Cloudy} simulation of the parameter grid, we set the ISM metal abundances to \texttt{abundances ISM}. We also include CMB heating and attenuation effects\footnote{Turning on the \texttt{CMB} in the \textsc{Cloudy} C17 setup makes the code report line luminosities after taking into account the isotropic CMB contrast by default, which is a non-negligible effect for high-$z$ [\ion{C}{2}] emission \citep[e.g.,][]{Lagache18,Kohandel19}. The attenuation effect is thus automatically included, unless it is manually disabled. See Section 7.1 of \cite{Ferland17}.}. The simulations are terminated at large distance ($\sim \,$kpc scale) from the incident surface, sufficient to interpolate $L_{\rm [CII],cl}$ of typical gas clouds.

From these \textsc{Cloudy} simulations of our chosen grid of parameters, we can learn how [\ion{C}{2}] emission depends on various conditions and environments of the multi-phase ISM, where we find that $\epsilon_{\rm [CII]}$ is more sensitive to $n_{\rm H}$ than other parameters (see more detailed discussion in Appendix~\ref{appendix:phase}). By assigning the interpolated $L_{\rm [CII],cl}$ to individual gas clouds, we obtain the distribution of [\ion{C}{2}] emission for a given simulated galaxy, which can then be projected onto a 2D plane to produce the image of [\ion{C}{2}] surface brightness ($\Sigma_{\rm [CII]}$). To enable direct comparison between UV continuum and [\ion{C}{2}] morphologies, we generate synthetic images of $\Sigma_{\rm [CII]}$ for each FIRE-2 galaxy at each snapshot, viewed along the same three sightlines defined in the \textsc{skirt} post-processing framework (Section~\ref{subsec:skirt}). In total, this produces 336 pairs of UV continuum and [\ion{C}{2}] images.

An example $\Sigma_{\rm [CII]}$ image is illustrated in Figure~\ref{fig:2d_images} (right panel). This example provides an initial view of how the extended [\ion{C}{2}] gas is distributed around young stars (left panel) and the UV continuum (middle panel). Further insights into the physical properties of [\ion{C}{2}]-emitting gas in simulated galaxies can be found in Figure~\ref{fig:phase}, which shows an example of the temperature ($T_{\rm gas}$) vs. density ($n_{\rm H}$) phase diagram. This phase diagram demonstrates that [\ion{C}{2}] arises predominantly in gas with $n_{\rm H} \sim 1-10^{3}\,{\rm cm^{-3}}$, in which scenario [\ion{C}{2}] is effectively optically thin \citep[e.g.,][]{Goldsmith12}, and that the FIRE-2 simulations are capable of properly resolving the phase structure of [\ion{C}{2}]-emitting gas.

\begin{figure*}[p]
\centering
\includegraphics[width=0.75\textwidth]{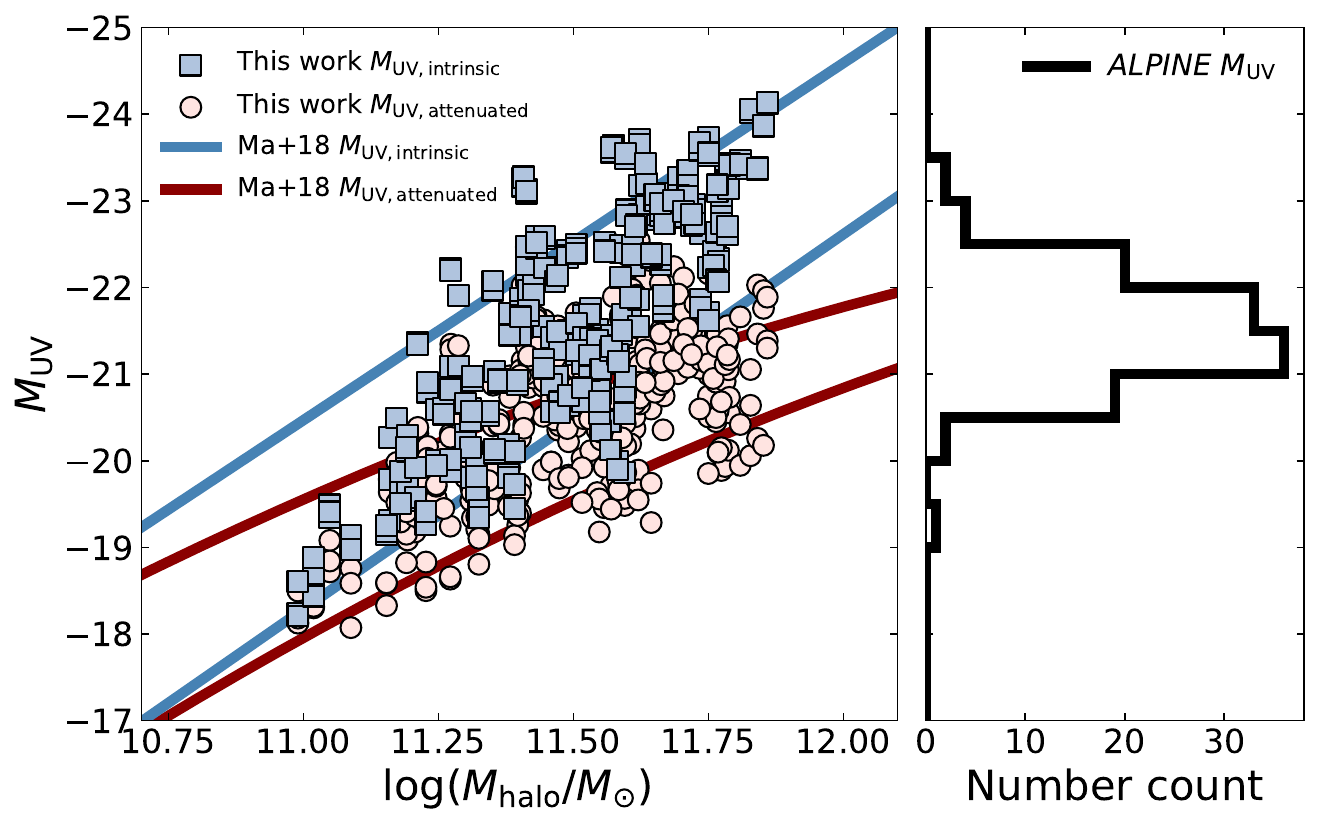}
\caption{Left panel: $M_{\mathrm{UV}}$--$M_{\mathrm{halo}}$ relation at $z \sim 6$ using the post-processed mock images of the UV continuum. The light blue and light red points indicate the intrinsic ($M_{\mathrm{UV,intrinsic}}$) and dust-attenuated ($M_{\mathrm{UV,attenuated}}$) UV magnitudes, respectively. We sample three viewing angles of the UV mock images for each snapshot of the simulated galaxies. Data points are compared with the $1\sigma$ range of scaling relations derived from \cite{Ma18}, where the red and blue curves represent $M_{\mathrm{UV,attenuated}}$ and $M_{\mathrm{UV,intrinsic}}$, respectively. Right panel: Distribution of the measured $M_{\mathrm{UV}}$ of ALPINE galaxies adopted from the main catalog in \cite{Faisst20}.
\label{fig:muv_mh}}
\end{figure*}

\begin{figure*}[h!]
\centering
\includegraphics[width=0.85\textwidth]{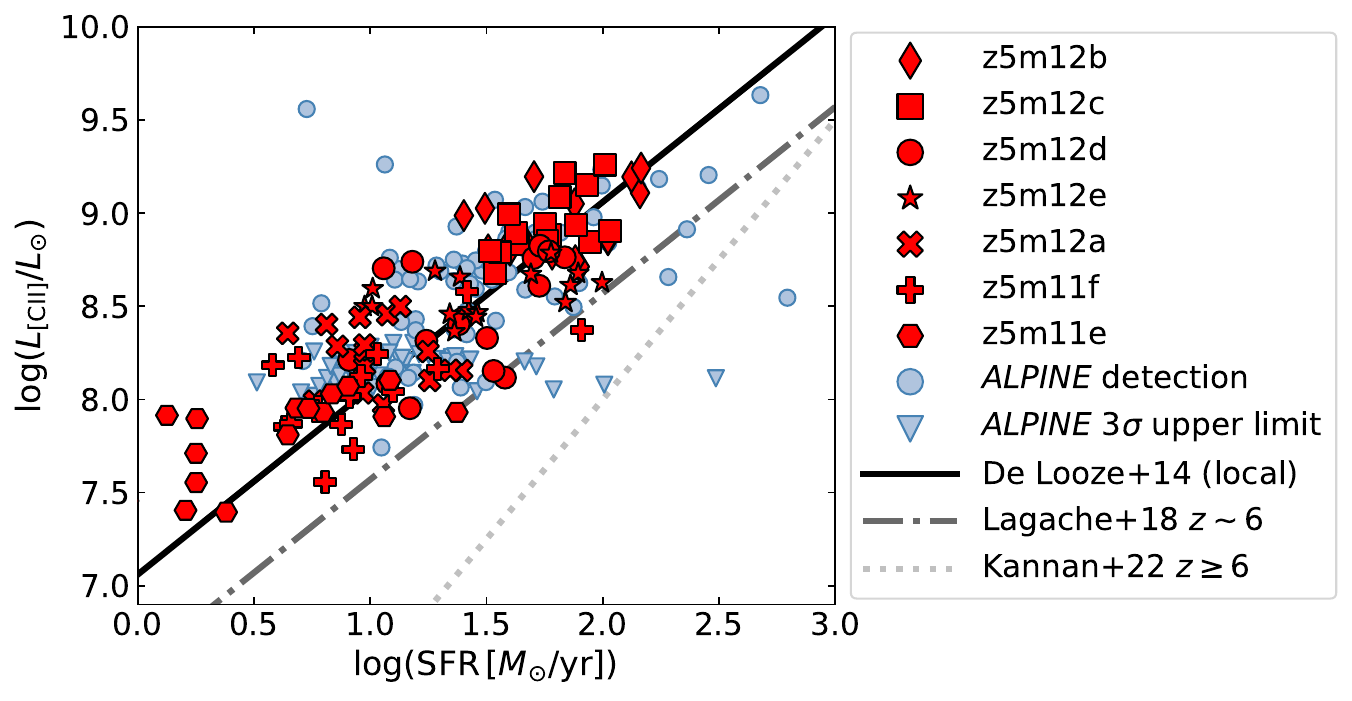}
\caption{Comparison between simulations and observations in the $L_{\mathrm{[CII]}}$--SFR plane. Red markers show our post-processing results of the seven FIRE-2 galaxies (each with 16 snapshots) over $5\leq z \leq6$. Blue data points show the measurements from the ALPINE survey, where circles are detections and triangles are $3\sigma$ upper limits \citep{Bethermin20,Faisst20,Schaerer20}. A few model predictions are shown for comparison: the black solid line shows the best-fit relation of local measurements \citep{Delooze14}, the gray dot-dashed line shows a semi-analytic model from \citet{Lagache18}, and the gray dotted line shows the best-fit relation from the \textsc{thesan} simulations \citep{Kannan22}.
}
\label{fig:cii_sfr}
\end{figure*}

\begin{figure*}[ht!]
\plotone{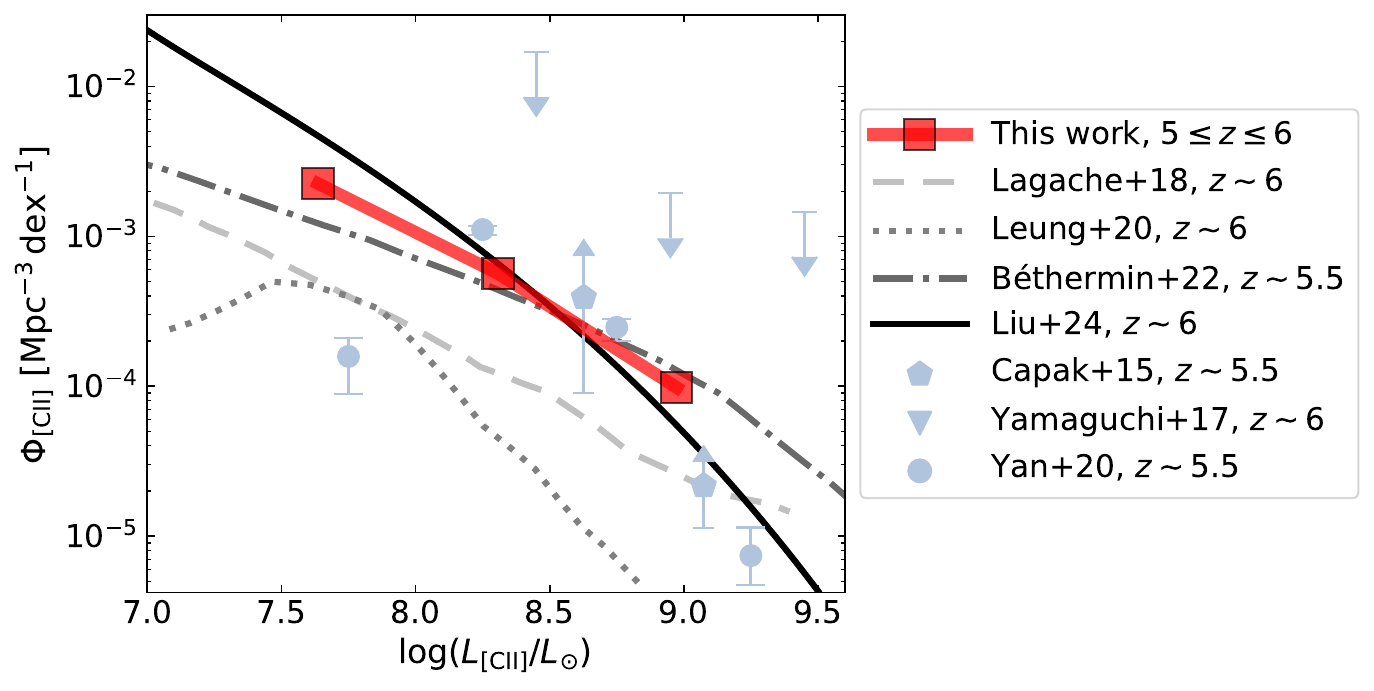}
\caption{Comparison between simulations and observations of the bright-end [\ion{C}{2}] luminosity function ($\Phi_{\mathrm{[CII]}}$) at $5 \leq z \leq 6$. Shown in red are predictions from our post-processing of FIRE-2 galaxies in three luminosity bins spanning $\log(L_{\rm [CII]}/L_{\odot}) = 7.3$--9.3. Blue data points and upper/lower limits are observational constraints \citep{Capak15,Yamaguchi17,Yan20}. We also show the comparison with several theoretical predictions: the solid black curve shows an analytic model that accounts for high-$z$ bursty star formation \citep{Liu24}, the dash-dotted curve \citep{Bethermin22} and dashed curve \citep{Lagache18} show predictions from semi-analytic models, and the dotted curve shows the results from the SIMBA simulations \citep{Leung20}.
\label{fig:ciilf}}
\end{figure*}

\section{Simulation Results: Comparison with Observations} \label{sec:compare}

\subsection{Galaxy Scaling Relations} \label{subsec:scaling}

We obtain the galaxy-integrated UV magnitudes, $M_{\rm UV}$, and [\ion{C}{2}] luminosities, $L_{\rm [CII]}$, of our selected galaxies using the post-processing framework. We examine these results through the $M_{\rm UV}$--$M_{\rm halo}$ and $L_{\rm [CII]}$--SFR relations, comparing them with observational data and other theoretical predictions to validate our approach and show that our results are representative.

Figure~\ref{fig:muv_mh} shows the $M_{\rm UV}$--$M_{\rm halo}$ relations with and without dust attenuation. The simulated results of the intrinsic UV magnitude ($M_{\rm UV,intrinsic}$) and the dust-attenuated UV magnitude ($M_{\rm UV,attenuated}$) of all simulated galaxies, redshift snapshots, and sightlines are displayed as scatter points. For the ease of comparing this relation with that investigated in \cite{Ma18}, we integrate the UV flux within $0.1 R_{\rm vir}$ ($\sim 3\,$kpc scale) of the simulated galaxies to obtain the galaxy-integrated $M_{\rm UV}$. Our simulation results of the entire sample are broadly consistent with the $\pm 1\sigma$ range of these scaling relations derived from \cite{Ma18}, which uses a simpler ray tracing method to calculate dust-attenuated $M_{\rm UV}$, in contrast to the Monte Carlo RT here. In the right panel of Figure~\ref{fig:muv_mh}, we show the distribution of $M_{\rm UV}$ measured at $1500\,\mathrm{\AA}$ for the entire ALPINE sample from the main catalog presented in \cite{Faisst20}, demonstrating that the simulated $M_{\rm UV}$ is comparable to observations and confirming that our selected FIRE-2 simulations are representative of the observed galaxy population.

Figure~\ref{fig:cii_sfr} shows the $L_{\mathrm{[CII]}}$--SFR relation of our simulated galaxies and its comparison with observations (the local relation from \citealt{Delooze14} and ALPINE $z\sim5$ galaxies from \citealt{Bethermin20,Faisst20,Schaerer20}) and other theoretical predictions \citep{Lagache18,Kannan22}. Taking into account the extended nature of [\ion{C}{2}], we sum $L_{\rm [CII],cl}$ of gas particles within $0.3R_{\rm vir}$ ($\sim 10\,$kpc scale, similar to the $1.5\arcsec$-radius aperture of ALPINE) of FIRE-2 galaxies to obtain the galaxy-integrated $L_{\rm [CII]}$, without applying a flux limit. The SFR of simulated galaxies is determined using the archaeological method that calculates the average SFR from stellar masses formed in the past $10\,$Myr, that is, ${\rm SFR} = {\rm SFR_{10Myr}}$, and we use it throughout this work (except for Section~\ref{subsec:3z} where ${\rm SFR_{10Myr}}$ is contrasted against ${\rm SFR_{100Myr}}$). The comparison between ALPINE galaxies and our simulations demonstrates good overall agreement, confirming that the results from our post-processing framework are representative of observations. Regarding outliers in the ALPINE sample that deviate from the linear trend, strong [\ion{C}{2}] deficits or non-detections may arise from one or multiple effects, such as low carbon abundance, intense radiation fields in compact starbursts, or observational factors like low surface brightness, though the exact causes remain unclear. Conversely, the two outliers significantly above the trend are associated with major mergers (see \citealt{HC25} and references therein). We also note that the fact that simulations such as \textsc{thesan} \citep{Kannan22} do not reproduce this scaling relation may reflect, at least in part, the importance of resolving the multiphase ISM structure, as in FIRE-2.

\subsection{Bright-End [\ion{C}{2}] Luminosity Function} \label{subsec:ciilf}

Apart from galaxy-integrated properties, it is useful to demonstrate that our simulations are representative and reasonable at the population level through the [\ion{C}{2}] luminosity function, $\Phi_{\mathrm{[CII]}}$. Given the range of selected FIRE-2 galaxies and redshift snapshots, we construct the bright end of $\Phi_{\mathrm{[CII]}}$ at $5 \leq z \leq 6$.

Following the method presented by \cite{Ma19}, we treat each snapshot as an independent `galaxy'. We collect all `galaxies' in a single redshift bin, $5 \leq z \leq 6$, where all 112 snapshots are used to sample the abundance of [\ion{C}{2}] emitters at the bright end. In each snapshot, a `galaxy' resides in a DM halo with $M_{\rm halo}$, and halo masses are categorized in three bins in the range of $\log(M_{\mathrm{halo}}/M_{\odot}) = 10.8-12$. The number count of sampled galaxies in the $i$-th $M_{\rm halo}$ bin, $N_{i}$, is subsequently linked with the expected abundance in the Universe by assigning a weight based on the halo mass function (HMF):
\begin{equation}
w_{i} = \phi_{i} \, \frac{\Delta\log M}{N_{i}},
\end{equation}
where $\phi_{i}$ is the HMF at the $M_{\rm halo}$ bin center and $\Delta\log M$ is the bin width. We use the public HMF\footnote{\url{https://hmf.readthedocs.io/en/latest/index.html}} code \citep{Murray13} and adopt the HMF model by \cite{Behroozi13} to calculate the HMF at $z = 5.5$. To obtain the bright-end $\Phi_{\mathrm{[CII]}}$, we categorize all galaxies into three $L_{\rm [CII]}$ bins in the $\log(L_{\mathrm{[CII]}}/L_{\odot}) = 7.3-9.3$ range and then calculate the number density of each $L_{\mathrm{[CII]}}$ bin by summing over their weights.

Figure~\ref{fig:ciilf} shows our prediction of the bright-end $\Phi_{\mathrm{[CII]}}$ at $5 \leq z \leq 6$ and the comparison with various models \citep{Lagache18,Leung20,Bethermin22,Liu24} and observational constraints \citep{Capak15,Yamaguchi17,Yan20} in the literature. The overall agreement demonstrates that our simulations and the post-processed galaxy sample are statistically representative. We note that in \cite{Yan20} the drop of $\Phi_{\mathrm{[CII]}}$ at $L_{\rm [CII]} \sim 10^{7.7} \, L_{\odot}$ might be due to selection effects associated with the UV-selected sample. At the faint end, line intensity mapping (LIM; see \citealt{Bernal22} for a recent review) provides a promising complementary approach. LIM measures the aggregate line emission from galaxies and large-scale structure without resolving individual sources, making it particularly sensitive to the faint-end behavior of $\Phi_{\mathrm{[CII]}}$ (e.g., \citealt{Sun2021,Liu24}, and references therein).

\subsection{Radial Profiles}\label{subsec:convolved}

\begin{figure}[h!]
\centering
\includegraphics[width=0.99\columnwidth]{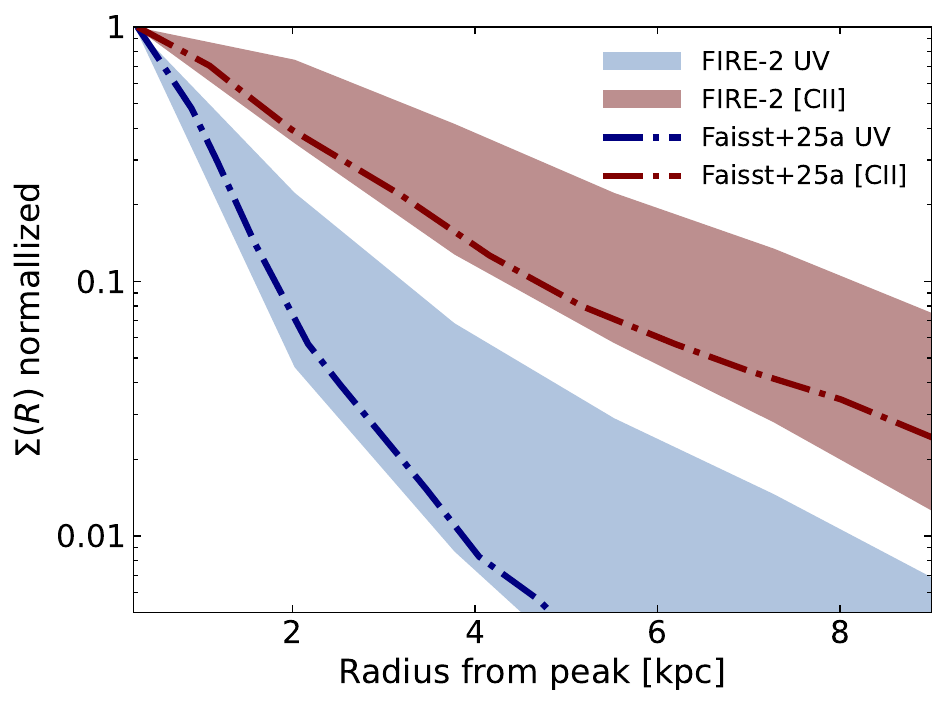}
\caption{Comparison of [\ion{C}{2}] and UV 1D profiles between our simulations and observations. The red and blue shaded bands show the 16--84th percentile range of simulated radial profiles of [\ion{C}{2}] and UV, respectively. The dash-dotted curves show the latest compilation of stacked profiles from the ALPINE-CRISTAL-JWST program \citep{Faisst2026a} for a subsample of isolated, non-merging galaxies comparable to our sample of simulated galaxies.
\label{fig:compare}}
\end{figure}

We investigate 1D radial profiles of [\ion{C}{2}] emission and UV continuum from our post-processing results and compare them with the latest observational constraints. Figure~\ref{fig:compare} displays these results. Our post-processing framework produces 336 pairs of [\ion{C}{2}] emission and UV continuum surface brightness images (7 FIRE-2 simulations $\times$ 16 redshift snapshots $\times$ 3 sightlines; see Section~\ref{sec:methods}). To compare with the latest observational constraints compiled by \cite{Faisst2026a}, we convolve these images with 2D Gaussian kernels with values of the full width at half-maximum (FWHM) similar to those of observations to simulate the effects of telescope beams. Although JWST and ALMA beams are not perfect 2D Gaussian profiles and exhibit variations, the Gaussian kernels provide a reasonable approximation for comparing simulated morphologies with observed radial profiles. Specifically, the UV continuum is convolved with a 2D Gaussian kernel with an $\mathrm{FWHM} = 0.1\,\arcsec$ ($\sim 0.6\,$kpc at $z = 5$; \citealt{Faisst2026a}), and the [\ion{C}{2}] emission is convolved with a 2D Gaussian with an $\mathrm{FWHM} = 0.3\,\arcsec$ ($\sim 2\,$kpc at $z = 5$, similar to the beam size of the CRISTAL program; \citealt{HC25,Faisst2026a}).

For each convolved image, we extract a 1D radial profile of average [\ion{C}{2}] and UV surface brightness in radial bins of width $1.5\,$kpc, comparable to observations. The center is defined by the peak UV continuum brightness ($\Sigma_{\rm UV,max}$) and is used for both [\ion{C}{2}] and UV radial profiles to facilitate comparisons between [\ion{C}{2}]-emitting gas and recent star-forming regions traced by the UV continuum. Nonetheless, we have verified that the choice of center (e.g., [\ion{C}{2}] peak vs. UV peak) does not significantly affect our results \citep[see also, e.g.,][]{Fujimoto19}. These profiles are normalized to their central surface brightness. We also exclude snapshots significantly contaminated by major mergers to focus our analysis on the [\ion{C}{2}] halo in central and satellite galaxies.

In Figure~\ref{fig:compare}, we summarize the normalized 1D profiles of $\Sigma_{\rm [CII]}$ and $\Sigma_{\rm UV}$ in shaded regions that encompass the 16th--84th percentiles of our simulation predictions. We compare these to the stacked average 1D profiles from a subsample of six isolated, non-merging galaxies from the latest ALPINE-CRISTAL-JWST observation \citep{Faisst2026a}. Notably, the selection of these galaxies aligns well with our simulation selection criteria that exclude major mergers. Our simulated results are broadly consistent with observations, demonstrating that the results of our framework can reproduce the concentrated UV continuum and extended [\ion{C}{2}] emission. In addition, the agreement in both the galaxy-integrated $L_{\rm [CII]}$ and the radial profile shapes implies that the normalization of the simulated emission is reasonably comparable to the observations. We further find the 16th--84th percentile contribution to the total $L_{\rm [CII]}$ from $3\,$kpc ($\sim 0.1 R_{\rm vir}$) and beyond (as a proxy for the CGM scale) to be $48$--$75\%$. The range of simulated UV profiles appears to be slightly more extended than the observed profile, which might be attributed to different beam profiles, the choice of sightlines, and strongly bursty star formation in FIRE-2 galaxies \citep[e.g.,][]{Sparre17,Stern21,Gurvich2023,Sun23b,Sun23}.

\begin{figure*}[ht!]
\centering
\includegraphics[width=0.99\textwidth]{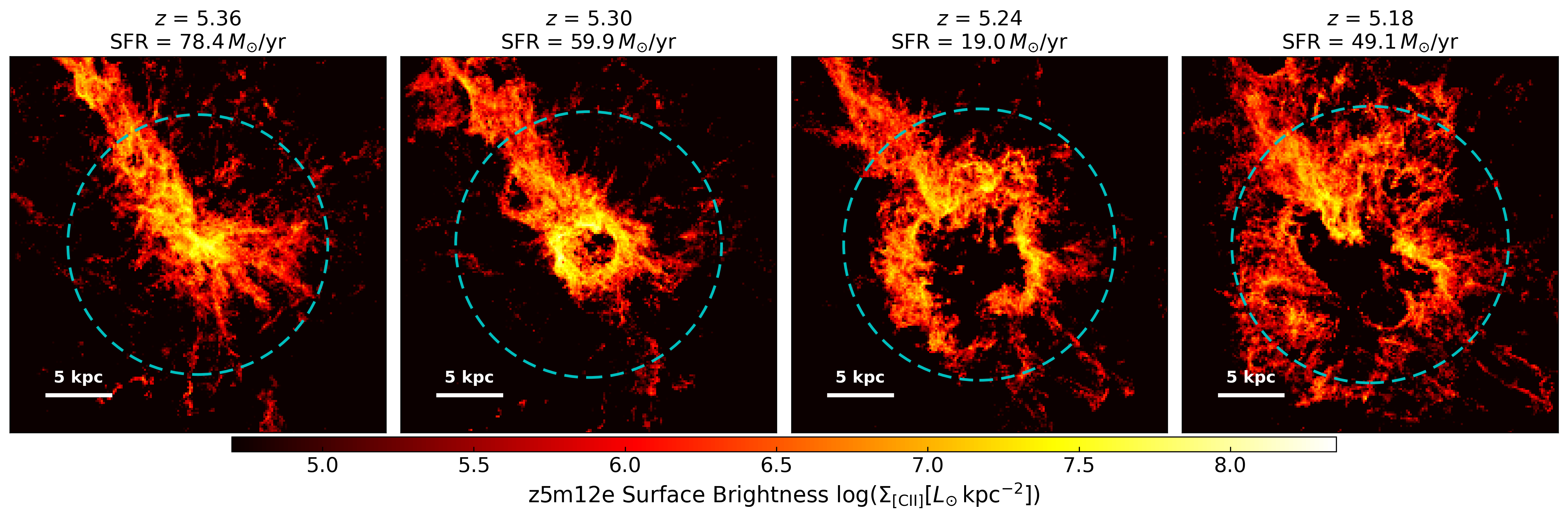}
\caption{Four consecutive redshift snapshots illustrating the evolution of the 2D [\ion{C}{2}] surface brightness of {\tt z5m12e} ($M_{\star} \sim 10^{10}\,M_{\odot}$ at $z = 5.3$) at a fixed viewing angle. These 2D maps show the raw simulated results without smoothing by a 2D Gaussian beam. From left to right, {\tt z5m12e} evolves from $z = 5.36$ to 5.18, with a time interval of $\sim15\,$Myr between every two consecutive snapshots. In each panel, the 2D distribution of [\ion{C}{2}] surface brightness is color-coded, the cyan dashed circle marks the reference scale for our [\ion{C}{2}] halo analysis ($0.3 R_{\rm vir}$ or $\sim 10\,$kpc), and the redshift and SFR are shown in the title. Note that the viewing angle is chosen to clearly illustrate the radial expansion of the [\ion{C}{2}]-emitting gas.
\label{fig:example_cii_halo}}
\end{figure*}

\section{Simulation Results: Evolution of Extended [\ion{C}{2}] Emission} \label{sec:explain}

\subsection{[\ion{C}{2}] Morphology Evolution} \label{subsec:z_evolution}

\begin{figure*}[p]
\centering
\includegraphics[width=0.99\textwidth]{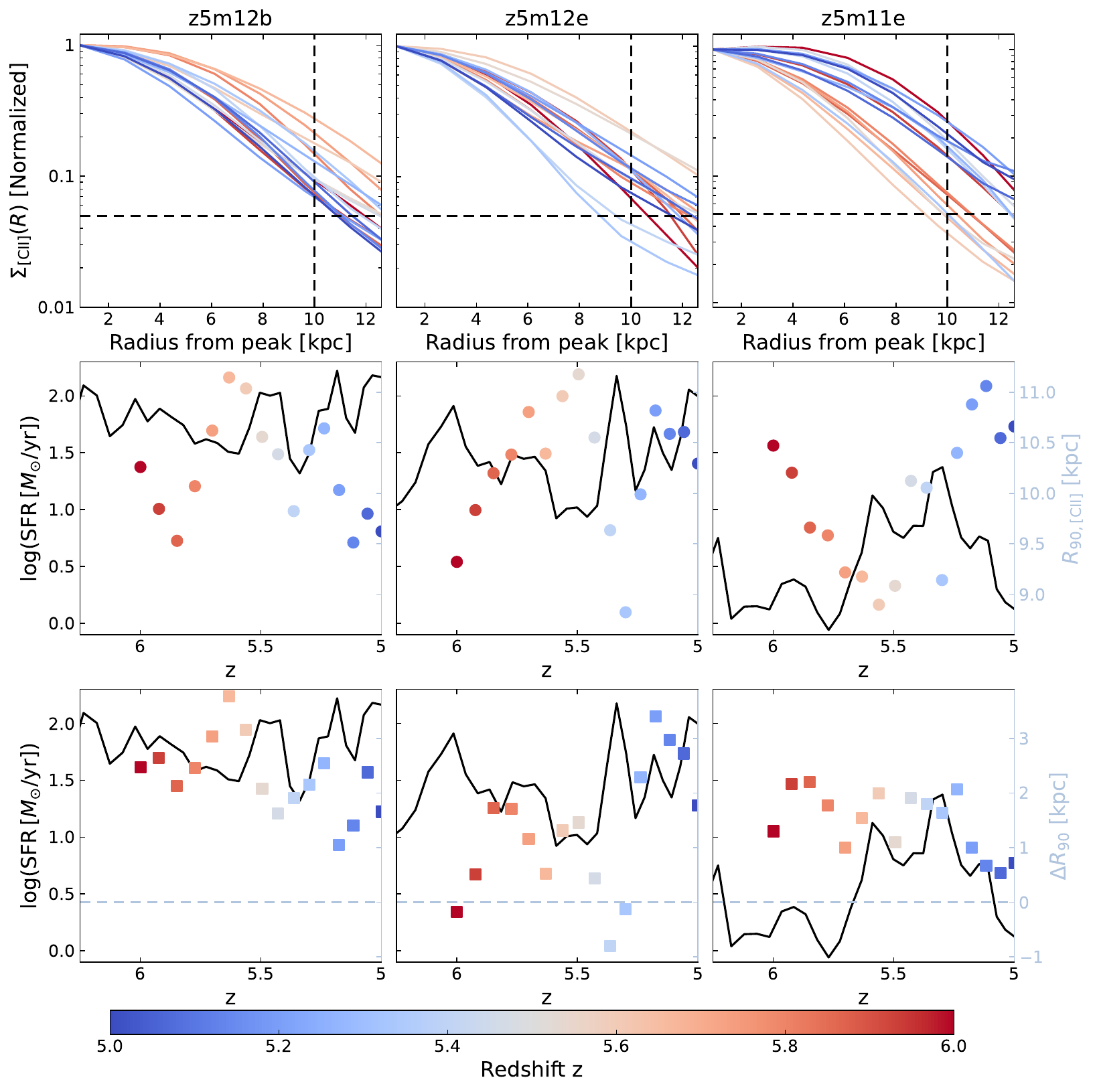}
\caption{Co-evolution of the size of [\ion{C}{2}] halos and the galaxy SFH. Three representative simulated galaxies in our sample are shown in the three columns, where \texttt{z5m12b} (left), \texttt{z5m12e} (middle), and \texttt{z5m11e} (right) represent our FIRE-2 galaxies in high ($M_{\star} \sim 10^{10.5}\,M_{\odot}$), medium ($M_{\star} \sim 10^{10}\,M_{\odot}$), and low ($M_{\star} \sim 10^{9.5}\,M_{\odot}$) mass ranges. The top row shows the smoothed, normalized 1D radial profile of [\ion{C}{2}] halos, color-coded by redshift. The cross point of the vertical and horizontal dashed lines shows one of the two requirements for an extended halo: a [\ion{C}{2}] profile should be sufficiently bright at $10\,$kpc, i.e., $\Sigma_{\rm [CII]}(10\,{\rm kpc}) \geq 0.05 \, \Sigma_{\rm [CII]}({\rm peak})$. The middle row shows each galaxy's $R_{90,\mathrm{[CII]}}$ in the context of its SFH (black solid), where we use $R_{90,\mathrm{[CII]}}$ to quantify the size of [\ion{C}{2}] halos and $R_{90,\mathrm{[CII]}}$ is the radius that contains $90\%$ of the total [\ion{C}{2}] flux. The redshifts are color-coded. The bottom row shows $\Delta R_{90} \equiv R_{90,\mathrm{[CII]}} - R_{90,\mathrm{UV}}$ in the context of SFHs (black solid), where for each redshift snapshot $\Delta R_{90} > 0$ indicates that the size of the [\ion{C}{2}] radial profile is larger than that of the UV continuum, in other words, [\ion{C}{2}] is more extended than UV and so can be labeled as an extended [\ion{C}{2}] halo.
\label{fig:cii_sfh}}
\end{figure*}

In Figure~\ref{fig:example_cii_halo}, we show raw (no beam smoothing) images corresponding to four consecutive snapshots of a FIRE-2 simulated galaxy ({\tt z5m12e}) as an example of how the morphology of [\ion{C}{2}] surface brightness evolves over a time interval of $\sim 50\,$Myr from $z = 5.36$ to 5.18. It is evident that the [\ion{C}{2}] morphology can evolve dramatically. At $z = 5.36$, [\ion{C}{2}] emission is centrally concentrated in the $R \lesssim 0.1 R_{\rm vir} \sim 3\,$kpc region, where the ISM is irradiated by recently formed stars. At $z = 5.18$, a substantial fraction of the [\ion{C}{2}]-emitting gas has migrated outward to inner-CGM scales ($0.3R_{\rm vir}$ or $\sim 10\,$kpc), suggesting that feedback from star formation activities plays a key role in driving [\ion{C}{2}]-emitting gas away from the center of the galaxy in this episode (see also \citealt{Sravan16} for a discussion of the bursty feedback producing time-variable UV metal line emission). The decrease in SFR during this episode indicates that this feedback simultaneously depletes the gas reservoir and suppresses star formation. The subsequent accretion and recycling of gas can replenish the reservoir, starting a new star formation episode and redistributing the [\ion{C}{2}] emission.

We note that the specific viewing angle in Figure~\ref{fig:example_cii_halo} (as well as Figure~\ref{fig:combined} in Appendix~\ref{appendix:more_snapshots}) is shown to optimally illustrate this radial expansion of the [\ion{C}{2}]-emitting gas in a straightforward manner. Some of the small-scale features will be substantially suppressed by the beam smoothing. We demonstrate this beam-smoothing effect in Figure~\ref{fig:smooth_combined} and discuss it further in Appendix~\ref{appendix:more_snapshots}.

\subsection{Bursty SFHs and the Evolution of [\ion{C}{2}] Halos} \label{subsec:bursty_sfh}

While visual inspection of 2D [\ion{C}{2}] morphology provides a qualitative understanding of [\ion{C}{2}] halo evolution, quantitative analysis of radial profiles is essential for (i) determining if a snapshot exhibits an extended [\ion{C}{2}] halo and (ii) self-consistently comparing [\ion{C}{2}] and UV radial profiles to study how [\ion{C}{2}] halos evolve in relation to the UV continuum and SFH.

To perform a quantitative investigation of [\ion{C}{2}] and UV radial profiles, we smooth both $\Sigma_{\rm [CII]}$ and $\Sigma_{\rm UV}$ images with the same 2D Gaussian kernel of ${\rm FWHM} = 1\arcsec$ ($\sim 6\,$kpc at $z = 5$), similar to the ALMA beam size used in \cite{Fujimoto19,Fujimoto20}. This kernel size, which is larger than the CRISTAL-like kernel size adopted in Section~\ref{subsec:convolved}, offers three key advantages. First, it produces smoothly varying radial profiles that suppress small-scale features and noise (compared to smaller beam sizes) that may complicate the classification of [\ion{C}{2}] halos. Second, it enables direct comparison between radial profiles of [\ion{C}{2}] emission and UV continuum by applying identical smoothing to both tracers. Third, it allows us to define extended [\ion{C}{2}] halos using observationally and physically motivated spatial scales.

Using this method, we generate [\ion{C}{2}] and UV radial profiles for all FIRE-2 simulated galaxies at $5 \leq z \leq 6$. Figure~\ref{fig:cii_sfh} shows the evolution of [\ion{C}{2}] halos of three representative FIRE-2 galaxies, {\tt z5m12b}, {\tt z5m12e}, and {\tt z5m11e}, representing the high ($M_{\star} \sim 10^{10.5}\,M_{\odot}$), medium ($M_{\star} \sim 10^{10}\,M_{\odot}$), and low ($M_{\star} \sim 10^{9.5}\,M_{\odot}$) mass ranges of the entire sample (see Table~\ref{tab:simtable}). As shown in this figure, we define that an extended [\ion{C}{2}] halo should meet the following two criteria:

\begin{itemize}
    \item The [\ion{C}{2}] emission should be sufficiently bright on the inner-CGM scales from the galaxy center, i.e., $\Sigma_{\mathrm{[CII]}}(10\,{\rm kpc}) \geq 0.05\Sigma_{\mathrm{[CII]}}({\rm peak})$.
    \item The 1D [\ion{C}{2}] profile of a snapshot should be more extended than the UV profile, i.e., $\Delta R_{90} > 0$.
\end{itemize}

Results of normalized [\ion{C}{2}] 1D profiles are shown in the top row of Figure~\ref{fig:cii_sfh}. The time variations are illustrated by the color coding of individual [\ion{C}{2}] profiles. In particular, the {\tt z5m12e} [\ion{C}{2}] profiles at $z \sim 5.3$--$5.4$ drop more rapidly than other profiles on the inner-CGM scales ($0.3 R_{\rm vir}$ or $\sim 10\,$kpc), indicating that the [\ion{C}{2}] emission at $z \sim 5.3$--$5.4$ is more concentrated, which confirms the information illustrated in Figure~\ref{fig:example_cii_halo}. This variation suggests that not all [\ion{C}{2}] profiles can be regarded as extended [\ion{C}{2}] halos, and a key aspect of extended [\ion{C}{2}] halos is that the [\ion{C}{2}] surface brightness $\Sigma_{\rm [CII]}$ ought to be sufficiently bright on inner-CGM scales. Therefore, we require $\Sigma_{\rm [CII]}(10\,{\rm kpc}) \geq 0.05 \, \Sigma_{\rm [CII]}({\rm peak})$, as indicated by dashed lines in Figure~\ref{fig:cii_sfh} (top row).

Based on 1D radial profiles of $\Sigma_{\rm [CII]}$ and $\Sigma_{\rm UV}$, we derive the radius enclosing 90\% of the total [\ion{C}{2}] ($R_{\rm 90,[CII]}$) and UV continuum ($R_{\rm 90,UV}$) emission as a parameter to quantify their spatial extent. We further define a metric $\Delta R_{90} \equiv R_{90,\mathrm{[CII]}} - R_{90,\mathrm{UV}}$ that quantifies the relative extent of [\ion{C}{2}] emission compared to UV continuum. In our analysis, $\Delta R_{90} > 0$ indicates that the [\ion{C}{2}] emission is more extended than the UV continuum, as shown by the horizontal dashed lines in the bottom row of Figure~\ref{fig:cii_sfh}.

The time evolution of $R_{\rm 90,[CII]}$ and $\Delta R_{90}$ in the context of SFHs during $5 \leq z \leq 6$ for three representative FIRE-2 galaxies is shown along the middle and bottom rows of Figure~\ref{fig:cii_sfh}, respectively. It is evident that these FIRE-2 galaxies exhibit highly bursty SFHs, a key feature of high-$z$ galaxies that has been predicted by both simulations and analytic arguments \citep[e.g.,][]{FM22,Mirocha23,Pallottini23,Shen23,Sun23b,Sun23,Liu24,Gelli24,Kravtsov24} and strongly supported by recent JWST observations \citep[e.g.,][]{Ciesla24,Looser24,Looser25,Cole25,Endsley25,Perry25,Munoz26}. SFHs of these FIRE-2 galaxies are characterized by multiple starburst cycles, each of which is an episode involving the rise, peak, and decline of the SFR. 

Examining the correlation between $R_{\rm 90,[CII]}$ or $\Delta R_{90}$ and the SFH during individual burst episodes, we find that the [\ion{C}{2}] emission often manifests as an extended [\ion{C}{2}] halo, and there appears to be a time-delayed correlation: The [\ion{C}{2}] halo size initially decreases as the SFR increases and approaches a local minimum near the SFR peak. Subsequently, as the SFR declines, the [\ion{C}{2}] emission expands, producing a [\ion{C}{2}] halo that is significantly more extended than the UV continuum. As shown in Figure~\ref{fig:cii_sfh}, both $R_{\rm 90,[CII]}$ and $\Delta R_{90}$ exhibit this trend, demonstrating that dramatic changes in the $R_{\rm 90}$ parameters are associated with recent starbursts, and thus extended [\ion{C}{2}] halos are strongly coupled to bursty SFHs. An outlier of this picture is the $z=6$ snapshot of {\tt z5m11e}, which is significantly contaminated by a nearby merger. Moreover, this qualitative pattern is confirmed by inspecting evolution tracks of $R_{\rm 90,[CII]}$ and $\Delta R_{90}$ vs. SFHs of other FIRE-2 galaxies, and $\Sigma_{\rm UV}$ and $\Sigma_{\rm [CII]}$ 2D images of all FIRE-2 galaxies (e.g., all {\tt z5m12e} snapshots in Figure~\ref{fig:combined}, Figure~\ref{fig:smooth_combined}, and Appendix~\ref{appendix:more_snapshots}).

We interpret this evolution as evidence that [\ion{C}{2}] halos are physically connected to bursty star formation and the associated stellar feedback. During a starburst, strong accretion and/or the recycling of previously expelled gas can feed the dense central ISM, making the [\ion{C}{2}] emission relatively more concentrated. Following the starburst, stellar feedback can disrupt the central gas reservoir and transport gas in physical conditions favorable for [\ion{C}{2}] emission to larger radii. As a result, the [\ion{C}{2}] morphology becomes more spatially extended during the post-burst phase. In this picture, different [\ion{C}{2}] morphologies reflect different phases of the same feedback-regulated baryon cycle, rather than tracing physically distinct galaxy populations.

\subsection{Properties and Occurrence Rate of [\ion{C}{2}] Halos} \label{subsec:3z}

The connection between the [\ion{C}{2}] halo size (measured by the $R_{90}$ parameters) and the SFH shown in Figure~\ref{fig:cii_sfh} hints at their potential correlation. To test whether such a correlation exists and more quantitatively describe the relationship between [\ion{C}{2}] halos and bursty star formation, we compute the temporal cross-correlation coefficient between the $R_{90}$ parameters and SFR as a function of the time lag $\Delta t$. Specifically, we calculate the Pearson correlation coefficients of $R_{90,{\rm [CII]}}(t)$ vs. ${\rm SFR}(t - \Delta t)$ and $\Delta R_{90}(t)$ vs. ${\rm SFR}(t - \Delta t)$ to investigate how [\ion{C}{2}] halo properties respond to SFR variations.

\begin{figure}[h!]
\centering
\includegraphics[width=0.99\columnwidth]{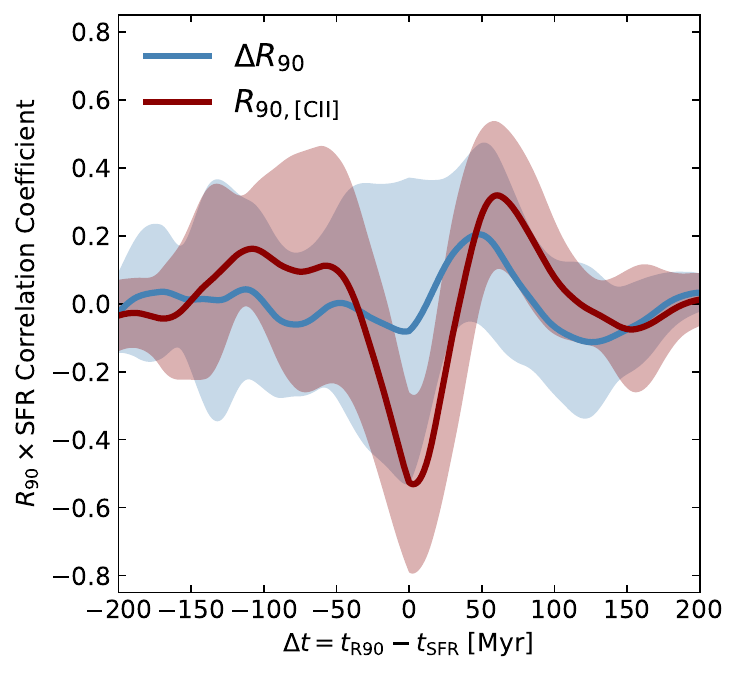}
\caption{Cross-correlation coefficient between the SFR and our two size measures of [\ion{C}{2}] halos ($R_{\rm 90,[CII]}$ and $\Delta R_{90}$), evaluated for all 21 evolution tracks (7 galaxies $\times$ 3 sightlines; each containing 16 snapshots over $z = 5$--6). The red curve shows the average correlation between the SFR and $R_{\rm 90,[CII]}$ as a function of the time lag $\Delta t$, whereas the blue curve shows the average correlation between the SFR and $\Delta R_{90}$. The shaded bands show the $\pm 1\sigma$ range. A clear negative correlation exists between the SFR and $R_{\rm 90,[CII]}$ at zero lag.}
\label{fig:corr_func}
\end{figure}

Figure~\ref{fig:corr_func} shows our findings. At zero time lag ($\Delta t \approx 0$), a significant negative correlation coefficient is present for the $R_{90,{\rm [CII]}}$--$\textrm{SFR}$ cross-correlation, which indicates that [\ion{C}{2}] halos tend to be more compact during starbursts when the SFR is high. The correlation becomes positive at a time lag of $\Delta t \sim 60\,$Myr, suggesting that larger [\ion{C}{2}] halos may be associated with a recent starburst $\sim 60\,$Myr earlier. This time-delayed positive correlation coefficient captures the physical process during which starburst-induced outflows drive [\ion{C}{2}]-emitting gas outward to inner-CGM scales within a characteristic timescale of $\sim 60\,$Myr. We also find no statistically significant correlation between $\Delta R_{90}$ and the ${\rm SFR}$, which may be associated with the attenuation of UV emission by dust, given its strong source-to-source and sightline-to-sightline variations.

This temporal correlation analysis implies an important connection between burst cycles and extended [\ion{C}{2}] halos: bursts of star formation power feedback that eventually creates spatially extended [\ion{C}{2}] emission, but the response is not instantaneous. Instead, [\ion{C}{2}]-emitting gas expands to large halo radii on a timescale of several tens of Myr following the recent starburst. This clearly demonstrates that the occurrence of extended [\ion{C}{2}] halos is linked to the recent SFH.

\begin{figure*}[p]
\centering
\includegraphics[width=0.49\textwidth]{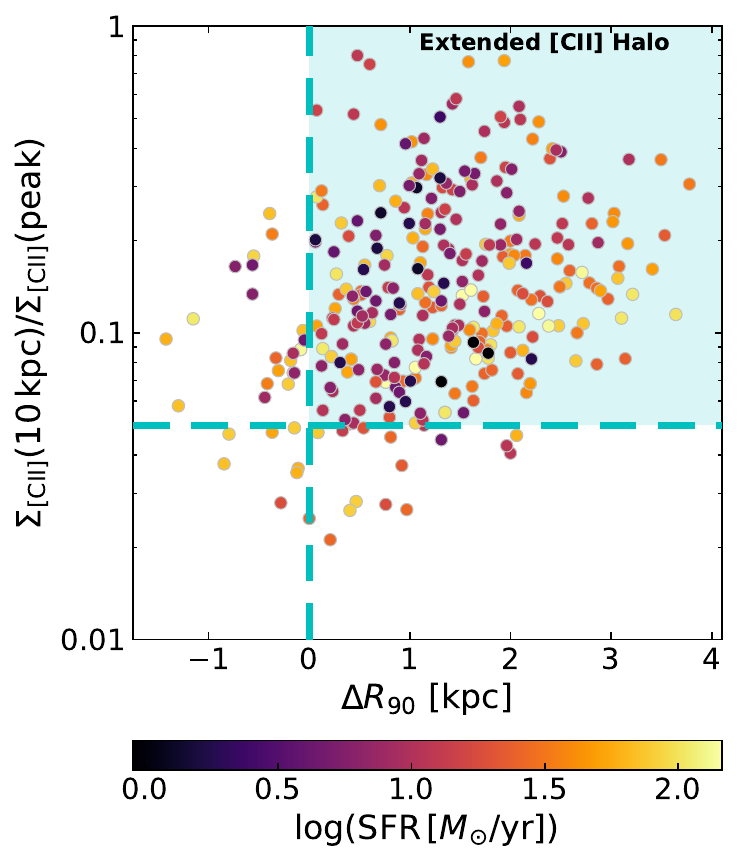}
\hfill
\includegraphics[width=0.49\textwidth]{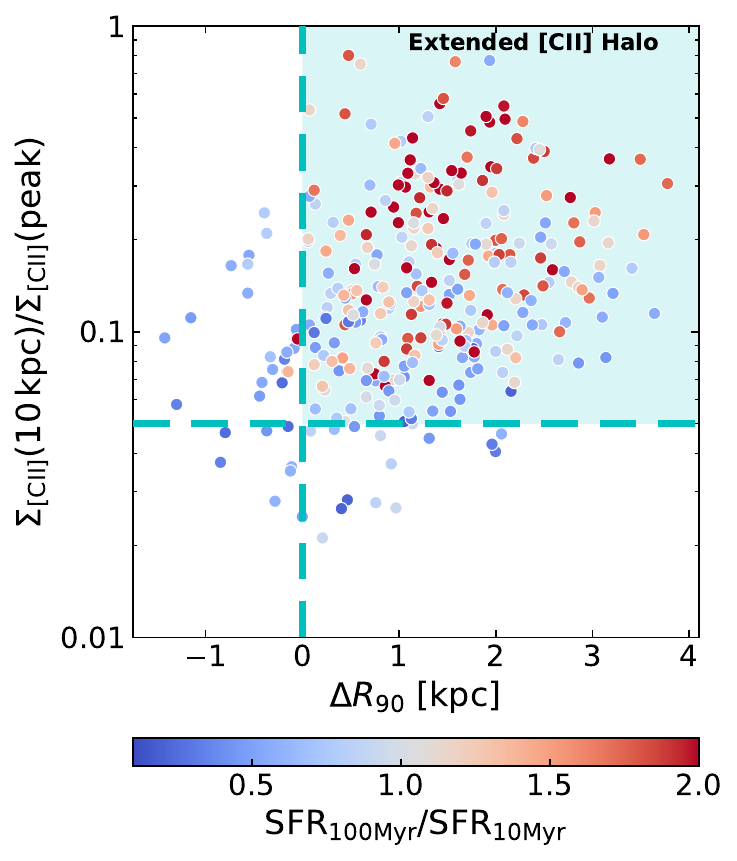}
\caption{Distribution of the simulated [\ion{C}{2}] profiles of all of our FIRE-2 galaxies (7 simulations $\times$ 16 snapshots $\times$ 3 sightlines) in the ($\Delta R_{90}$, $\Sigma_{\rm [CII]}(10\,{\rm kpc})/\Sigma_{\rm [CII]}({\rm peak})$) plane. Dashed lines show the criteria to label the extended [\ion{C}{2}] halo: (i) $\Delta R_{90} > 0$ and (ii) $\Sigma_{\rm [CII]}(10\,{\rm kpc})/\Sigma_{\rm [CII]}({\rm peak}) > 0.05$. As such, data points of extended [\ion{C}{2}] halo (upper right) can be distinguished from concentrated [\ion{C}{2}] emission (all others). Left: The color-coded SFR of each snapshot represents the average SFR of the recent $10\,$Myr (${\rm SFR_{10Myr}}$). Right: The color-coded SFR of each snapshot is the ratio between the average SFR of the recent $100\,$Myr (${\rm SFR_{100Myr}}$) and ${\rm SFR_{10Myr}}$, i.e., ${\rm SFR_{100Myr}/SFR_{10Myr}}$.
\label{fig:scatter}}
\end{figure*}

We further plot the statistical distribution of [\ion{C}{2}] emission of all 336 snapshots in Figure~\ref{fig:scatter} based on our classification criteria (cyan dashed lines): (i) $\Sigma_{\mathrm{[CII]}}(10\,{\rm kpc}) \geq 0.05\Sigma_{\mathrm{[CII]}}({\rm peak})$ and (ii) $\Delta R_{90} > 0$. We find that the occurrence rate of the extended [\ion{C}{2}] halo equals $86\%$ in our $5 \leq z \leq 6$ FIRE-2 galaxies. We also note that \cite{Fujimoto20} reported that the occurrence rate of [\ion{C}{2}] halos in isolated ALPINE galaxies is $\sim60\%$\footnote{Recent follow-up observations at higher spatial resolution have revealed that some sources previously classified as isolated in earlier studies are actually merging systems \citep{Posses25}. This might slightly affect the fraction of isolated galaxies reported in \cite{Fujimoto20}.}. While our simulations suggest a slightly higher occurrence rate, this may be attributed to multiple factors, such as differences in classification criteria and sample selection.

\begin{figure*}[ht!]
\centering
\includegraphics[width=0.99\textwidth]{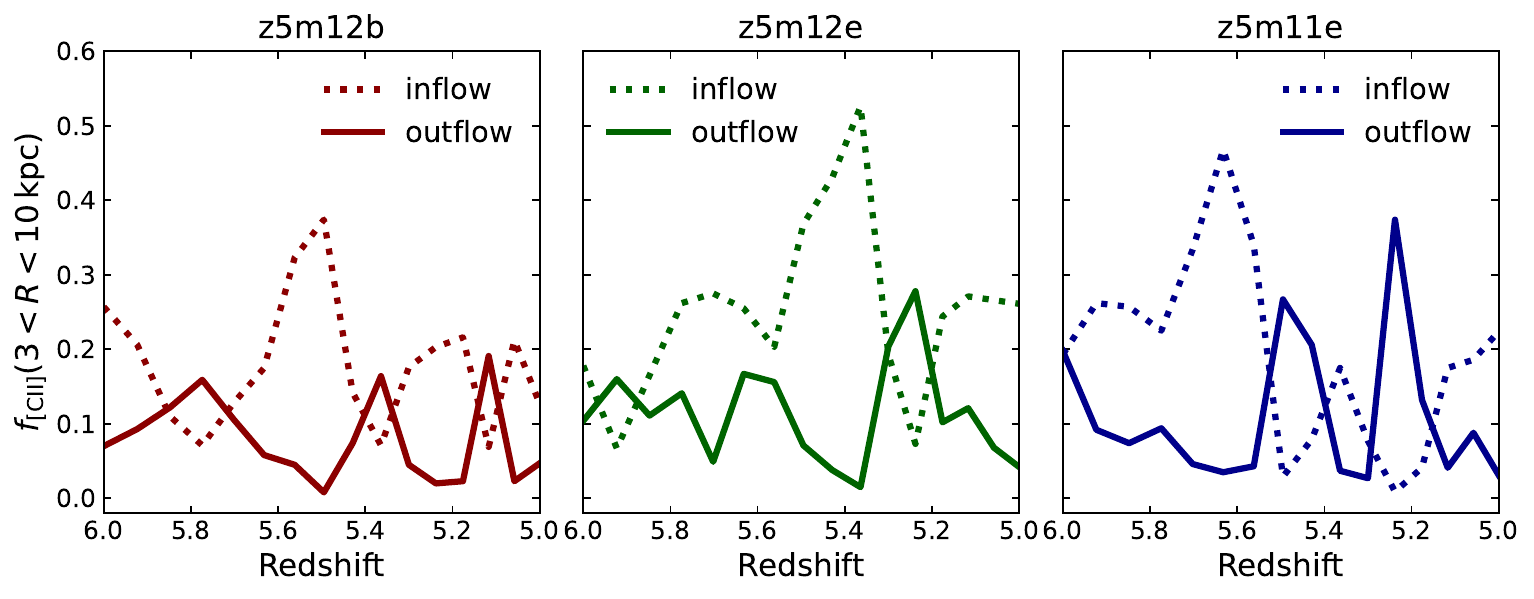}
\caption{Time evolution of the fractional contribution to the total [\ion{C}{2}] luminosity by outflowing gas (solid curves) and inflowing gas (dotted curves) in three representative simulated galaxies in our sample, where {\tt z5m12b} (left panel), {\tt z5m12e} (middle panel), and {\tt z5m11e} (right panel) represent our FIRE-2 galaxies in high ($M_{\star} \sim 10^{10.5}\,M_{\odot}$), medium ($M_{\star} \sim 10^{10}\,M_{\odot}$), and low ($M_{\star} \sim 10^{9.5}\,M_{\odot}$) mass ranges. A gas particle of a galaxy is classified as inflowing/outflowing by examining its radial velocity pointing toward/away from the center of mass of a DM halo with a magnitude greater than the velocity dispersion. This fractional contribution is studied in the CGM scales of $3 < R < 10\,$kpc and from $z=6$ to 5.
\label{fig:vr_frac}}
\end{figure*}

Moreover, data points in Figure~\ref{fig:scatter} are color-coded by SFR$_\mathrm{10\,Myr}$ (left) and SFR$_\mathrm{100\,Myr}$/SFR$_\mathrm{10\,Myr}$ (right). From the left panel, there appears to be no clear trend between the [\ion{C}{2}] halo size and SFR$_\mathrm{10\,Myr}$, whereas from the right panel, the ratio ${\rm SFR_{100Myr}/SFR_{10Myr}}$ characterizes recent SFH variations. ${\rm SFR_{10Myr}}$ is sensitive to the formation of massive stars and is typically estimated from tracers such as H$\alpha$, as it directly probes ionizing photons from OB stars with lifetimes $\lesssim 10$--$20\,$Myr. On the other hand, ${\rm SFR_{100Myr}}$ reflects the cumulative star formation over longer timescales and is typically estimated from the UV continuum \citep[e.g.,][]{Sparre17,FV21}. Particularly, ${\rm SFR_{100Myr}/SFR_{10Myr}} < 1$ indicates a recently rising SFH, whereas ${\rm SFR_{100Myr}/SFR_{10Myr}} > 1$ suggests a recent decline in the SFH. This ratio is therefore fundamental for quantifying SFH variability, identifying burst/quenching phases, and understanding the drivers of galaxy evolution on sub-Gyr timescales. Indeed, the ${\rm UV}/{\rm H\alpha}$ luminosity ratio has been commonly used in observations to investigate bursty star formation in high-$z$ galaxies \citep[e.g.,][]{Faisst19,Mehta23,Cole25,Looser25,Munoz26}.

We find that nearly all ${\rm SFR_{100Myr}/SFR_{10Myr}} > 1$ data points can be classified as [\ion{C}{2}] halos by our criteria. This corroborates the idea that extended [\ion{C}{2}] emission correlates with the SFH---a starburst in the recent past is associated with the emergence of extended [\ion{C}{2}] halos, as has been shown in Figure~\ref{fig:cii_sfh}. We also find that the ${\rm SFR_{100Myr}/SFR_{10Myr}} < 1$ data points are widely scattered across the plotted space, though more extreme recent starbursts shown as deep blue points tend to exhibit concentrated [\ion{C}{2}] emission.

\subsection{[\ion{C}{2}] from Inflows, Outflows, and Turbulent Gas}\label{subsec:gas_dispersion}

With the strong correlation between [\ion{C}{2}] halos and starburst episodes revealed, it is crucial to understand the kinematics of [\ion{C}{2}]-emitting gas and investigate the role of outflows and inflows in producing extended [\ion{C}{2}] emission. The direction of gas radial motion can be determined by the radial component of the velocity vector $v_{\rm gas,rad} \hat{r}$ with respect to the center of mass. We note that it may not be sufficient to use the sign of $v_{\rm gas,rad} \hat{r}$ to classify a gas particle as outflowing/inflowing \citep[as in, e.g.,][]{Munoz24}, given that some non-negligible fraction of the gas particles may be consistent with the intrinsic randomness in the gas motion, instead of coherent gas flows moving outward/inward. This is particularly true for the low-mass, bursty galaxies of interest here, whose gas reservoir is primarily supported by turbulent rather than rotational motion \citep[e.g.,][]{Stern21,Gurvich2023,Sun26}. Therefore, in our analysis, a gas particle is regarded as outflowing/inflowing only when its radial velocity is greater than the $1\sigma$ velocity dispersion in magnitude \citep{Muratov15}, and as consistent with random motion if its radial velocity falls within the $1\sigma$ velocity dispersion. We calculate these fractions using all [\ion{C}{2}]-emitting gas without applying a flux limit. We can then investigate the fractional contribution, $f_{\rm [CII]} \equiv L_{\rm [CII]}({\rm outflow \,\, or \,\, inflow})/L_{\rm [CII],tot}$, of gas to the [\ion{C}{2}] emission on inner-CGM scales ($3 < R < 10\,$kpc) of our simulated galaxies.

Figure~\ref{fig:vr_frac} shows the evolution of $f_{\rm [CII]} (3 < R < 10\,{\rm kpc})$ for the same example FIRE-2 galaxies as in Figure~\ref{fig:cii_sfh}. Overall, outflows and inflows account for a modest fraction of the inner-CGM scale [\ion{C}{2}]-emitting gas, with the sum of their fractional contributions, $f_{\rm [CII],outflow}+f_{\rm [CII],inflow}$, typically remaining below $\sim 50\%$. The remaining [\ion{C}{2}] emission resides in gas particles without statistically significant radial motion beyond the characteristic velocity dispersion. Notably, $f_{\rm [CII],outflow}$ and $f_{\rm [CII],inflow}$ exhibit a clear anti-correlation: when outflow contributions increase, inflow contributions decrease, and vice versa. The temporal evolution of these components is correlated with the [\ion{C}{2}] halo evolution and starburst episodes. During the initial phase of a starburst, the inflow contribution typically increases as gas accretes to fuel star formation. Following the starburst peak, the [\ion{C}{2}] halo becomes more extended ($R_{\rm 90,[CII]}$ and $\Delta R_{\rm 90}$ increase) as the outflow contribution rises. This outflowing gas may subsequently be recycled to replenish the ISM and initiate the next starburst episode if not gravitationally unbound.

\subsection{[\ion{C}{2}] from Gravitationally Unbound Gas}\label{subsec:gas_escape}

\begin{figure}[h!]
\centering
\includegraphics[width=0.99\columnwidth]{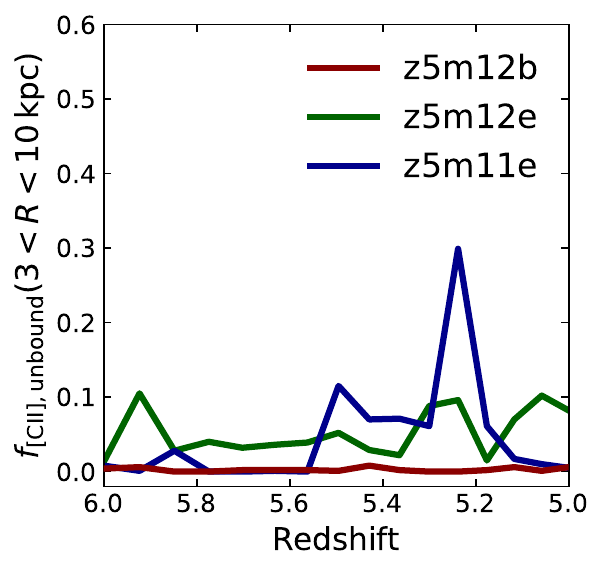}
\caption{Time evolution of the fractional contribution to the total [\ion{C}{2}] luminosity by gravitationally unbound gas ($f_{\rm [CII],unbound} \equiv L_{\rm [CII],unbound}/L_{\rm [CII],tot}$, with $v \geq v_{\rm escape}$) in three examples of our FIRE-2 simulations, {\tt z5m12b} (red), {\tt z5m12e} (green), and {\tt z5m11e} (blue), which represent the high ($M_{\star} \sim 10^{10.5}\,M_{\odot}$), medium ($M_{\star} \sim 10^{10}\,M_{\odot}$), and low ($M_{\star} \sim 10^{9.5}\,M_{\odot}$) mass ranges. In {\tt z5m12b}, {\tt z5m12e}, and {\tt z5m11e}, $v_{\rm escape} \sim 549, 442, 334\,$km/s respectively. This fractional contribution is studied on inner-CGM scales of $3 < R < 10\,$kpc from $z=6$ to 5.}
\label{fig:unbound}
\end{figure}

Given that starburst-driven outflows can expel [\ion{C}{2}]-emitting gas to larger distances from the galaxy center, a natural question arises: Can these outflows achieve sufficiently high velocities to be gravitationally unbound? To address this, we characterize the kinematics of inner-CGM scale [\ion{C}{2}]-emitting gas by examining the gravitationally bound vs. unbound fractions. We classify a gas particle as unbound if its velocity exceeds the local escape velocity ($v > v_{\rm escape}$), that is, it should have sufficient kinetic energy to move from inner-CGM scales ($3<R<10\,$kpc) to $R_{\rm vir}$ of the host DM halo. Since DM dominates the total mass, the $v_{\rm escape}$ is determined primarily by the DM halo's gravitational potential.

Given that DM halos of our simulated galaxies are well described by the Navarro-Frenk-White (NFW) profile \citep{Wetzel23,Wetzel25}, we calculate the gravitational potential of an NFW profile with a concentration $c\sim 4$ and determine the $v_{\rm escape}$ threshold for each of our FIRE-2 simulations. Then, we define $f_{\rm [CII],unbound} \equiv L_{\rm [CII],unbound}/L_{\rm [CII],tot}$ as the fraction of [\ion{C}{2}] luminosity contributed by unbound gas within the $3 < R < 10\,$kpc inner-CGM scale. Figure~\ref{fig:unbound} shows the evolution of $f_{\rm [CII],unbound}(3<R< 10\,{\rm kpc})$ for the same example galaxies as in Figure~\ref{fig:cii_sfh}.

Despite the presence of starburst cycles, the majority of [\ion{C}{2}]-emitting gas on inner-CGM scales remains gravitationally bound. For example, {\tt z5m12e} is less massive than {\tt z5m12b}, and its $f_{\rm [CII],unbound}$ can sometimes achieve $\sim 10\%$. {\tt z5m11e} is in the low-mass regime of our sample, whose $f_{\rm [CII],unbound}$ can sometimes become $\sim 30\%$. We note that the spike at $z \sim 5.2$ for {\tt z5m11e} can be found in both Figure~\ref{fig:vr_frac} and Figure~\ref{fig:unbound}, suggesting that outflows can sometimes drive gas to high kinetic energies to become gravitationally unbound.

\subsection{[\ion{C}{2}] from Satellite Galaxies}\label{subsec:satellite}

Extended [\ion{C}{2}] halos may also be contributed by [\ion{C}{2}] emission from satellite galaxies orbiting the central galaxy. \cite{Fujimoto19,Fujimoto20} first proposed this scenario, and recent cosmological simulations have shown that satellites can produce significant extended [\ion{C}{2}] emission around massive galaxies \citep{Munoz24,Khatri25}.

\begin{figure}[h!]
\centering
\includegraphics[width=0.99\columnwidth]{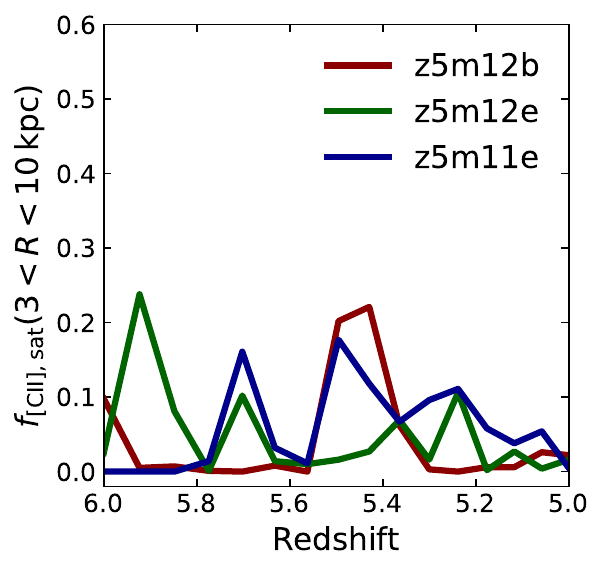}
\caption{Time evolution of the fractional contribution to the total [\ion{C}{2}] luminosity by satellite galaxies that reside in subhalos ($f_{\rm [CII],sat} \equiv L_{\rm [CII],sat}/L_{\rm [CII],tot}$) in three examples of our FIRE-2 simulations, {\tt z5m12b} (red), {\tt z5m12e} (green), and {\tt z5m11e} (blue), which represent high ($M_{\star} \sim 10^{10.5}\,M_{\odot}$), medium ($M_{\star} \sim 10^{10}\,M_{\odot}$), and low ($M_{\star} \sim 10^{9.5}\,M_{\odot}$) mass ranges. This fractional contribution is studied in the inner-CGM scales of $3 < R < 10\,$kpc and from $z=6$ to 5.
\label{fig:sat_frac}}
\end{figure}

To quantify the fractional contribution of satellites to [\ion{C}{2}] halos, we examine [\ion{C}{2}]-emitting gas associated with subhalos that host satellite galaxies. In practice, we identify all subhalos (i) located in the range of $0.1R_{\rm vir} < R < R_{\rm vir}$ ($0.1R_{\rm vir}$ corresponds to the $\sim 3\,$kpc scale) of the central DM halo of the simulation, and (ii) with masses in the range $10^{8}\,M_{\odot} \leq M_{\rm subhalo} \leq 0.2M_{\rm halo}$, where $M_{\rm halo}$ denotes the mass of the central halo excluding subhalos. This mass range ensures that the selected subhalos are massive enough to host star-forming satellites while excluding galaxies in subhalos with mass $M_{\rm subhalo} > 0.2M_{\rm halo}$, which can be classified as major mergers. For each identified subhalo, we then calculate the [\ion{C}{2}] luminosity within $0.3R_{\rm vir,subhalo}$ of the subhalo center and constrain their contribution to the total inner-CGM scale $L_{\rm [CII]}$, defined as $f_{\rm [CII],sat} \equiv L_{\rm [CII],sat}/L_{\rm [CII],tot}$.

Figure~\ref{fig:sat_frac} shows the evolution of $f_{\rm [CII],sat}(3<R< 10\,{\rm kpc})$ for the same example FIRE-2 galaxies as Figure~\ref{fig:cii_sfh}. It is evident that satellite galaxies provide a modest contribution to [\ion{C}{2}] halos in early galaxies, with $f_{\rm [CII],sat}$ typically remaining below $\sim 20\%$. In addition, we note that the satellite contribution to the $z \sim 5.9$ snapshot of {\tt z5m12e} can be clearly seen in Figure~\ref{fig:combined}, in which a bright satellite galaxy dominates $f_{\rm [CII],sat}$ on inner-CGM scales.

\section{Discussion} \label{sec:discussion}

\subsection{[\ion{C}{2}] Halos in Various Theoretical Studies} \label{subsec:compare}

In this work, we have modeled the spatial distribution of [\ion{C}{2}] emission in high-$z$ galaxies using the FIRE-2 cosmological zoom-in simulations. Our post-processing framework successfully reproduces the ubiquitous extended [\ion{C}{2}] halos observed out to the inner CGM ($\sim 10\,$kpc) in high-$z$ galaxies. We have investigated the temporal evolution of [\ion{C}{2}] halos and revealed a strong correlation between the [\ion{C}{2}] halo size and bursty SFH. Compared to several previous theoretical investigations of extended [\ion{C}{2}] emission, our approach introduces a few notable improvements, which we discuss below.

First, compared to semi-analytical models, which rely on parameterized prescriptions and scaling relations, cosmological simulations provide more realistic and comprehensive representations of the environments and evolution in high-$z$ galaxies. For example, \cite{Pizzati20,Pizzati23} developed a semi-analytical model focused on the outflow scenario, demonstrating that outflows can produce extended [\ion{C}{2}] emission. However, through detailed treatment of physical conditions and processes in FIRE-2 galaxies with resolved ISM, we find that outflows are not the only cause of [\ion{C}{2}] halos on the inner-CGM scale---inflows and turbulent gas also play important roles (Section~\ref{subsec:gas_dispersion}).

Second, given that [\ion{C}{2}] originates from multiple gas phases, it is essential for simulations to resolve the phase structure of [\ion{C}{2}]-emitting gas. The FIRE-2 simulations achieve sub-pc resolution in the densest gas (Section~\ref{subsec:fire2}). As demonstrated in Section~\ref{subsec:Cloudy}, the FIRE-2 simulations properly capture the phase structure of [\ion{C}{2}]-emitting gas. In contrast, most previous simulation-based studies of [\ion{C}{2}] halos do not achieve such high resolution required to fully resolve the multi-phase origin of [\ion{C}{2}] emission. For example, \cite{Munoz24} use TNG50 galaxies with $\sim 100\,$pc resolution, which rely on an assumed effective equation of state to describe the multi-phase ISM. This lack of adequate resolution may cause these simulations to miss some key aspects needed to accurately model [\ion{C}{2}] emission. Additionally, different feedback prescriptions may impact the spatial extent of [\ion{C}{2}] emission, given our demonstrated correlation between [\ion{C}{2}] halos and starburst episodes. Weaker feedback strength can result in more compact [\ion{C}{2}] distributions, potentially explaining why some prior studies underpredict [\ion{C}{2}] halo sizes \citep[e.g.,][]{Pallottini17,Pallottini19,Pallottini22,Schimek24}. This may also connect to the CGM contribution to total $L_{\rm [CII]}$: for example, \cite{Schimek24} found $\sim10\%$ of [\ion{C}{2}] emission originates from the CGM, whereas we typically find $\gtrsim 50\%$ (Section~\ref{subsec:convolved}). Moreover, we study the temporal evolution of spatially resolved [\ion{C}{2}] emission in relation to the SFH (see Section~\ref{subsec:bursty_sfh}), an aspect not addressed in previous work.

Lastly, our analysis employs carefully defined metrics and well-motivated classifications for [\ion{C}{2}] halos. (i) We introduce the $\Delta R_{90}$ parameter, which quantifies how much more extended [\ion{C}{2}] emission is compared to UV continuum, following the observations \citep[e.g.,][]{Fujimoto19,Fujimoto20}. While recent simulations \citep{Munoz24,Khatri25} partially reproduce extended [\ion{C}{2}] emission, they did not compare with the UV continuum profile for complete consistency with the observations. (ii) We more strictly define outflows/inflows as gas with radial velocity greater than the velocity dispersion, which can distinguish outflows/inflows with respect to the random motion along the radial direction. This contrasts with \cite{Munoz24}, which classified gas solely by the sign of radial velocity. The simpler approach by \cite{Munoz24} may significantly overestimate the outflowing/inflowing gas contribution by neglecting the strong random motion of gas in high-$z$ bursty galaxies. (iii) Different definitions of satellite galaxies can lead to varying estimates of their contribution to [\ion{C}{2}] emission on the inner-CGM scale. We define satellites as subhalos with $M_{\rm subhalo} \leq 0.2 M_{\rm halo}$ and measure their [\ion{C}{2}] emission within $0.3 R_{\rm vir,subhalo}$ from each subhalo center, yielding a modest satellite contribution with $f_{\rm [CII],sat}$ typically below $\sim 20\%$ (Section~\ref{subsec:satellite}). This contrasts with, e.g., \cite{Khatri25}, who found that satellites are the primary driver of extended [\ion{C}{2}] emission.

\subsection{Observability of [\ion{C}{2}] Halos at Higher Redshift} \label{subsec:z7}
 
With an increasing number of $z \gtrsim 7$ galaxies being observed by ALMA \citep[e.g.,][]{Bouwens22}, including confirmations of JWST-detected galaxies \citep[e.g.,][]{Zavala24}, statistical samples of [\ion{C}{2}] halos around individual galaxies may soon be assembled at higher redshift, motivating questions such as: Do the prevalence and size of [\ion{C}{2}] halos evolve? Here, we further predict the [\ion{C}{2}] halos in $6.5 \leq z \leq 7.5$ snapshots of our FIRE-2 galaxies.

We select snapshots in our FIRE-2 galaxies with $M_{\star}$ comparable to those from the ALMA-REBELS survey \citep{Bouwens22}. This selection gives {\tt z5m12b} and {\tt z5m12c}, the most massive simulated galaxies in our sample run down to $z = 5$, whose simulated $M_{\star}$ ranges from $10^{8.7}$ to $10^{9.9} M_{\odot}$ at $6.5 \leq z \leq 7.5$ (comparable to the mass range $10^{8.5} M_{\odot} \lesssim M_{\rm \star} \lesssim 10^{10.5} M_{\odot}$ of the ALMA-REBELS sample). We obtain a sample of 60 snapshots that comprises 2 galaxies $\times$ 10 snapshots $\times$ 3 sightlines. We post-process these galaxies following the same procedure mentioned above to extract 1D radial profiles from smoothed images.

To study the detectability of these radial profiles, we derive observed fluxes in units of ${\rm mJy}$, following \cite{Solomon92}:
\begin{equation}
S_{\rm [CII]} = \frac{10^{6}}{1.04}\frac{L_{\rm [CII]}}{\nu_{\rm obs} \, D_{\rm L}^{2} \, \Delta v},
\end{equation}
where $L_{\rm [CII]}$ is in units of $L_{\odot}$, $\nu_{\rm obs}$ is the observed frequency of [\ion{C}{2}] emission in units of ${\rm GHz}$, $D_{\rm L}$ is the luminosity distance in units of ${\rm Mpc}$, and $\Delta v$ is the velocity range of observed spectra, for which we adopt $\Delta v = 300 \, {\rm km/s}$ common for high-$z$ galaxies \citep[e.g.,][]{Bethermin20,HC25}. We then convert the flux to surface density (in units of ${\rm mJy/arcsec^{2}}$).

\begin{figure}[h!]
\centering
\includegraphics[width=0.99\columnwidth]{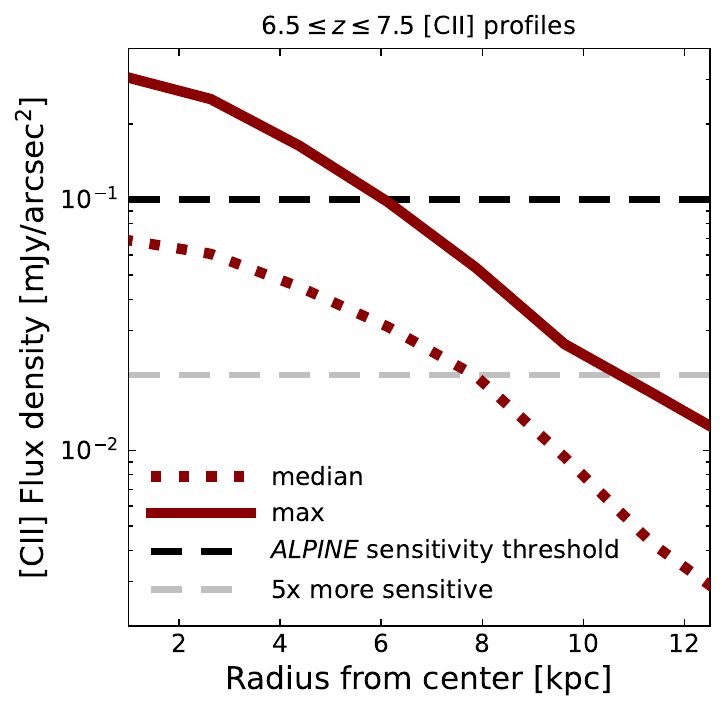}
\caption{Predicted [\ion{C}{2}] profile of $z \sim 7$ massive galaxies and its detectability by ALMA. Red curves plot [\ion{C}{2}] radial profiles extracted from the most massive simulated galaxies in our sample ({\tt z5m12b} and {\tt z5m12c}) at $6.5\leq z \leq 7.5$. These are typical massive galaxies at $z\sim7$ \citep[e.g.,][]{Bouwens22}. The solid red curve shows the profile of the brightest possible [\ion{C}{2}], and the dotted red curve shows the median profile. To predict the detectability of these $z\sim7$ extended [\ion{C}{2}] halos, we plot horizontal dashed lines which are simple estimates of the $5\sigma$ [\ion{C}{2}] line sensitivity threshold based on the ALPINE survey \citep{Bethermin20,Fujimoto20}. The black dashed line shows the sensitivity threshold similar to that of the ALPINE survey, and the gray line shows the level of $5\times$ better sensitivity.
\label{fig:z7_predict}}
\end{figure}

Figure~\ref{fig:z7_predict} shows the simulated results of $z \sim 7$ [\ion{C}{2}] radial profiles and their comparison with a simple estimate of $5\sigma$ sensitivity thresholds based on the ALPINE survey \citep{Bethermin20,Fujimoto20}. We extract the brightest and median [\ion{C}{2}] profiles by taking the maximum and median values at each radial bin across all snapshots. We find that even the brightest [\ion{C}{2}] profile cannot be detected given the ALPINE sensitivity. For a sensitivity threshold $5\times$ better than ALPINE, it becomes possible to detect the brightest $\sim 10\,$kpc [\ion{C}{2}] halo profiles in our sample. For reference, the most massive $z \sim 7$ galaxy in our sample has $M_{\star} = 10^{9.9} \, M_{\odot}$, ${\rm SFR} = 40\, M_{\odot}\,{\rm yr}^{-1}$, and $L_{\rm [CII]} = 10^{8.6} \, L_{\odot}$. We stress that to maximize the chance of detecting [\ion{C}{2}] halos at $z \sim 7$, target galaxies should be more massive, have higher SFR, or be more [\ion{C}{2}] bright than this reference galaxy. Using a $5\times$ more sensitive detection threshold typically results in a $25\times$ longer integration time. On the other hand, the [\ion{C}{2}] line at $z \sim 7$ shifts from ALMA Band 7 (for ALPINE galaxies) to Band 6. Under comparable weather conditions, the improved atmospheric transmission and lower system temperature in Band 6 reduce the integration time by $\sim 40\%$ compared to Band 7. Given that the average observation time for an ALPINE galaxy is $\sim20\,$minutes \citep{Bethermin20,Fujimoto20}, we may require $\sim 5$ hours of on-source time on ALMA to achieve a $5\times$ better sensitivity to measure the $\sim 10\,$kpc [\ion{C}{2}] halo in $z \sim 7$ massive star-forming galaxies. With its future updates including the Wideband Sensitivity Upgrade \citep{Carpenter23}, ALMA's increased spectral imaging speed will significantly reduce the required integration time.

\subsection{Limitations and Future Work} \label{subsec:future}

Here we note several limitations of this work and discuss several avenues for improvements and future studies. Regarding the FIRE simulations, the new FIRE-3 model includes updated stellar evolution tracks, star formation criteria, low-temperature cooling and chemistry, and black hole physics \citep{Hopkins23}. These changes could affect the sub-kpc (or even $\sim 10 \,$pc) scale properties of high-$z$ galaxies. We defer the comparison between [\ion{C}{2}] halos in FIRE-2 and FIRE-3 simulations to future work. Our sample is limited to massive galaxies ($M_{\star} \gtrsim 10^{9}\,M_{\odot}$) at $5 \leq z \leq 6$. Extending this analysis to lower-mass galaxies would provide further insights into the nature of [\ion{C}{2}] halos and bursty star formation. We defer to future work for a detailed analysis of [\ion{C}{2}] halos in a wide range of high-$z$ galaxy populations. In addition, the FIRE-2 simulations do not include black hole physics and AGN feedback (Section~\ref{subsec:fire2}), whereas recent observations have revealed evidence of AGN activity in a few ALPINE galaxies previously classified as star-forming galaxies \citep[e.g.,][]{Barchiesi23,Fujimoto25}. We emphasize that these simulations offer a clean, controlled sample for the majority of ALPINE-like galaxies, which are likely primarily regulated by stellar feedback, and we defer an investigation of the impact of AGN on [\ion{C}{2}] halos to future work.

Regarding the post-processing framework, parameters such as dust properties are calibrated to the Milky Way, which could deviate from high-$z$ galaxies. We note that recent work has shown that the MW calibration is still valid, such as the MW extinction curve \citep{Ferrara22}, suggesting that the change in calibration will not significantly affect our post-processing results. When modeling the dust radiative transfer using \textsc{skirt}, we mainly consider the ISRF from star particles; therefore, the metagalactic UV background \citep[e.g.,][]{Faucher20} is not included in our [\ion{C}{2}] halo analysis. The UV background is subdominant to the local ISRF on the scales of interest in our FIRE-2 galaxies, while this background might become relevant at larger distances from the galaxy center ($> 10\,$kpc). When simulating observed [\ion{C}{2}] emission in \textsc{Cloudy}, while in this framework \textsc{Cloudy} takes a single number of $J_{\rm FUV}$, we note that the photoionization can also be affected by the SED shape. Since we just find small variations of SED shape in $6-13.6\,$eV, we consider it as a minor effect that does not significantly affect the \textsc{Cloudy} post-processing results. In addition, some state-of-the-art simulations now incorporate on-the-fly RT and non-equilibrium thermochemistry to accurately predict emission line diagnostics of high-$z$ galaxies \citep[e.g.,][]{Katz22,Choustikov2026}, which serve as a useful cross check and can motivate modeling assumptions involved in the post-processing analysis.

We note that while the gas kinematics is mainly regulated by accretion and stellar feedback, it can be further complicated by mergers and galaxy interactions, where tidal stripping and ram pressure may also play significant roles. In our FIRE-2 sample, we have found a few snapshots that suggest that the extended [\ion{C}{2}] emission might also be tied to major merger events, but with only a few instances, we do not have enough information to reliably characterize this connection. A detailed investigation of how mergers and galaxy interactions shape gas kinematics and [\ion{C}{2}] halos is beyond the scope of this paper (but see, e.g., \citealt{DC24} for recent simulations of extended [\ion{C}{2}] emission in merging systems).

It should be noted that the framework used in this work is by no means unique to high-$z$ [\ion{C}{2}] emission. Other essential emission lines that trace galaxy evolution, such as Ly$\alpha$, [\ion{O}{3}]$_{\rm \lambda\lambda4959,5007}$, H$\alpha$, and [\ion{O}{3}]$\,88\,\mu$m, can also be simulated on FIRE-2 galaxies to perform morphological investigations. Particularly, \cite{Stern21} found that Ly$\alpha$ can be used to trace neutral gas in the inner CGM, given high optical depth and rapid cooling, while this work demonstrates that the [\ion{C}{2}] halo traces the enrichment processes in the inner CGM. Recent observations also showed the extended nature of [\ion{O}{3}]$_{\rm \lambda\lambda4959,5007}$ and H$\alpha$ \citep[e.g.,][]{Fischer18,Faisst2026a,Trefoloni25,Wang26}. Moreover, the [\ion{O}{3}]$\,88\,\mu$m line is a primary target for forthcoming ALMA observation programs that aim to probe dust-obscured star formation in $z \gtrsim 8$ galaxies, such as the PHOENIX program (ALMA\#2025.1.01606.L, PI: Schouws; in conjunction with UV observations by JWST GO\#9425, PI: Schouws). It is compelling to incorporate these lines in the post-processing analysis in future work.

In addition, if the extended [\ion{C}{2}] halo is prevalent among high-$z$ galaxies, this might impact the [\ion{C}{2}] LIM signals in the one-halo regime. Recently, \cite{Zhang23} studied the response of the [\ion{C}{2}] power spectrum to variations of [\ion{C}{2}] halo sizes. It would be interesting to investigate the population-level constraints of [\ion{C}{2}] halos from the one-halo [\ion{C}{2}] LIM signal in the FIRE simulations.

\section{Conclusions} \label{sec:conclusion}

In this paper, we explore the origin and evolution of extended [\ion{C}{2}] emission that prevails in high-$z$ bursty galaxies. We implement a post-processing framework on a sample of $5 \leq z \leq 6$ galaxies ($M_{\star} \sim 10^{9.5}$--$10^{10.5}\,M_{\odot}$) from the FIRE-2 cosmological zoom-in simulations, which involves simulating the three-dimensional dust radiative transfer using \textsc{skirt} and photoionization modeling using \textsc{Cloudy}. Our post-processing analysis not only enables direct comparisons with ALMA observations, but also allows us to investigate in detail how [\ion{C}{2}] halos evolve in response to the bursty SFHs of high-$z$ galaxies. 

Our main findings are summarized as follows:

\begin{itemize}
    \item The predictions of our post-processing framework reproduce the observed galaxy scaling relations and statistics at high redshifts, including the $M_{\mathrm{UV}}$--$M_{\mathrm{halo}}$ and $L_{\mathrm{[CII]}}$--SFR relations, along with the bright-end $\Phi_{\mathrm{[CII]}}$.
    \item We generate synthetic images and extract 1D radial profiles of [\ion{C}{2}] emission and UV continuum, from which we find broad agreement between simulated profiles and observations. In particular, a significant portion (typically $\gtrsim 50\%$) of the [\ion{C}{2}] emission comes from the inner CGM, and thus [\ion{C}{2}] is often more spatially extended than UV.
    \item Based on our classification criteria, we find that the rate of occurrence of extended [\ion{C}{2}] halos is $\gtrsim 80\%$. This result is broadly consistent with observations and confirms the widespread emergence of [\ion{C}{2}] halos on inner-CGM scales ($\sim 10\,$kpc).
    \item By analyzing the evolution of extended [\ion{C}{2}] halos in the context of bursty SFHs, we find a strong correlation between the [\ion{C}{2}] halo size and starburst episodes: the [\ion{C}{2}] halo size decreases while the central star-forming region accretes gas and boosts SFR; the [\ion{C}{2}] halo size significantly increases after the starburst peak, a process that expels [\ion{C}{2}]-emitting gas to the inner CGM and suppresses SFR; [\ion{C}{2}]-emitting gas can later be recycled to restart the next starburst episode. This is further quantified by the correlation coefficient between the time evolution of [\ion{C}{2}] halo size and the bursty SFH, from which we find that the former correlates with the latter with a characteristic time delay of $\sim 60\,$Myr, likely due to the outflowing of [\ion{C}{2}]-emitting gas driven by a recent starburst.
    \item By studying the [\ion{C}{2}] halo size in the context of the ${\rm SFR_{100Myr}/SFR_{10Myr}}$ ratio, which traces variability of the recent SFH, we find that starburst-induced outflows drive the [\ion{C}{2}] emission to be more extended. This provides additional support for the connection between [\ion{C}{2}] halo size and the bursty SFH. Complementary H$\alpha$ observations, together with [\ion{C}{2}] and UV continuum morphologies, would be essential for further testing this prediction.
    \item Using a stringent definition of gas outflow/inflow, we find that outflows and inflows contribute modestly to the inner-CGM-scale [\ion{C}{2}] emission. Their contribution is strongly correlated with starburst episodes. We also find that most of the inner-CGM [\ion{C}{2}]-emitting gas remains gravitationally bound.
    \item The contribution of satellite galaxies to the [\ion{C}{2}] halo is modest, with the fractional contribution typically remaining below $\sim20\%$.
    \item We predict that it is possible to measure inner-CGM-scale [\ion{C}{2}] halos at $z \sim 7$ with ALMA. Measuring the [\ion{C}{2}] halo of a single massive star-forming galaxy at $z \sim 7$ may require $\sim 5$ hours of on-source integration time on ALMA.
\end{itemize}

In summary, this study reveals the correlation between the extended [\ion{C}{2}] halo and bursty SFH in high-$z$ galaxies, and highlights the complex nature of [\ion{C}{2}] halos. It opens an avenue for using the spatial information of line emission to probe star formation activity and the phase of starburst episodes in early galaxies. It is compelling to perform detailed investigations of the impact of gas kinematics on [\ion{C}{2}] halos and observations. Future observations will be able to measure [\ion{C}{2}] halos in a wider mass and redshift range of high-$z$ objects, which can help us better understand the interplay of star formation, the ISM, and the CGM. Moreover, the post-processing framework presented is by no means limited to modeling [\ion{C}{2}]. It can be applied to other important emission line diagnostics, such as H$\alpha$ and [\ion{O}{3}], which can potentially offer further insight into high-$z$ galactic ecosystems.
 
\begin{acknowledgments}

We thank the anonymous referee for insightful comments and suggestions, which have improved the clarity and quality of this manuscript. We thank Caleb Choban, Phil Hopkins, Cameron Hummels, Matt Bradford, and Tzu-Ching Chang for helpful discussions and comments during the preparation of this paper. L.J.L. acknowledges Jonas Zmuidzinas for his help with the NRAO SOS proposal, which supports this work through the NSF NRAO award SOSPADA-037. G.S. was supported by a CIERA Postdoctoral Fellowship. C.A.F.G. was supported by NSF through grants AST-2108230 and AST-2307327; by NASA through grants 21-ATP21-0036 and 23-ATP23-0008; and by STScI through grant JWST-AR-03252.001-A. AL acknowledges support from BSF Grant \# 2024193.

The authors use simulations from the FIRE-2 public data release \citep{Wetzel23,Wetzel25}. The FIRE-2 cosmological zoom-in simulations of galaxy formation are part of the Feedback In Realistic Environments (FIRE) project, generated using the \textsc{gizmo} code \citep{Hopkins15} and the FIRE-2 physics model \citep{Hopkins18}. The authors acknowledge the Texas Advanced Computing Center (TACC) at The University of Texas at Austin for providing computational resources that have contributed to the research results reported within this paper. URL: http://www.tacc.utexas.edu. This paper makes use of the following ALMA data: ADS/JAO.ALMA\#2017.1.00428.L and 2021.1.00280.L. ALMA is a partnership of ESO (representing its member states), NSF (USA) and NINS (Japan), together with NRC (Canada), NSTC and ASIAA (Taiwan), and KASI (Republic of Korea), in cooperation with the Republic of Chile. The Joint ALMA Observatory is operated by ESO, AUI/NRAO and NAOJ. The National Radio Astronomy Observatory is a facility of the National Science Foundation operated under cooperative agreement by Associated Universities, Inc. This work is based on observations taken by the 3D-HST Treasury Program (GO 12177 and 12328) with the NASA/ESA HST, which is operated by the Association of Universities for Research in Astronomy, Inc., under NASA contract NAS5-26555.

\end{acknowledgments}

\facilities{ALMA, HST, TACC}

\software{astropy \citep{Astropy13,Astropy18,Astropy22},  
          \textsc{Cloudy} \citep{Ferland17}, 
          HMF \citep{Murray13}, 
          matplotlib \citep{Hunter07}, 
          NumPy \citep{Harris20}, 
          pandas \citep{pandas10,pandas20}, 
          sklearn \citep{Pedregosa11}, 
          SciPy \citep{Virtanen20},
          \textsc{skirt} \citep{Camps15}
          }

\appendix

\section{Parameter Dependence of [\ion{C}{2}] Emission in \textsc{Cloudy}} \label{appendix:phase}

\begin{figure*}[ht!]
\centering
\includegraphics[width=0.99\textwidth]{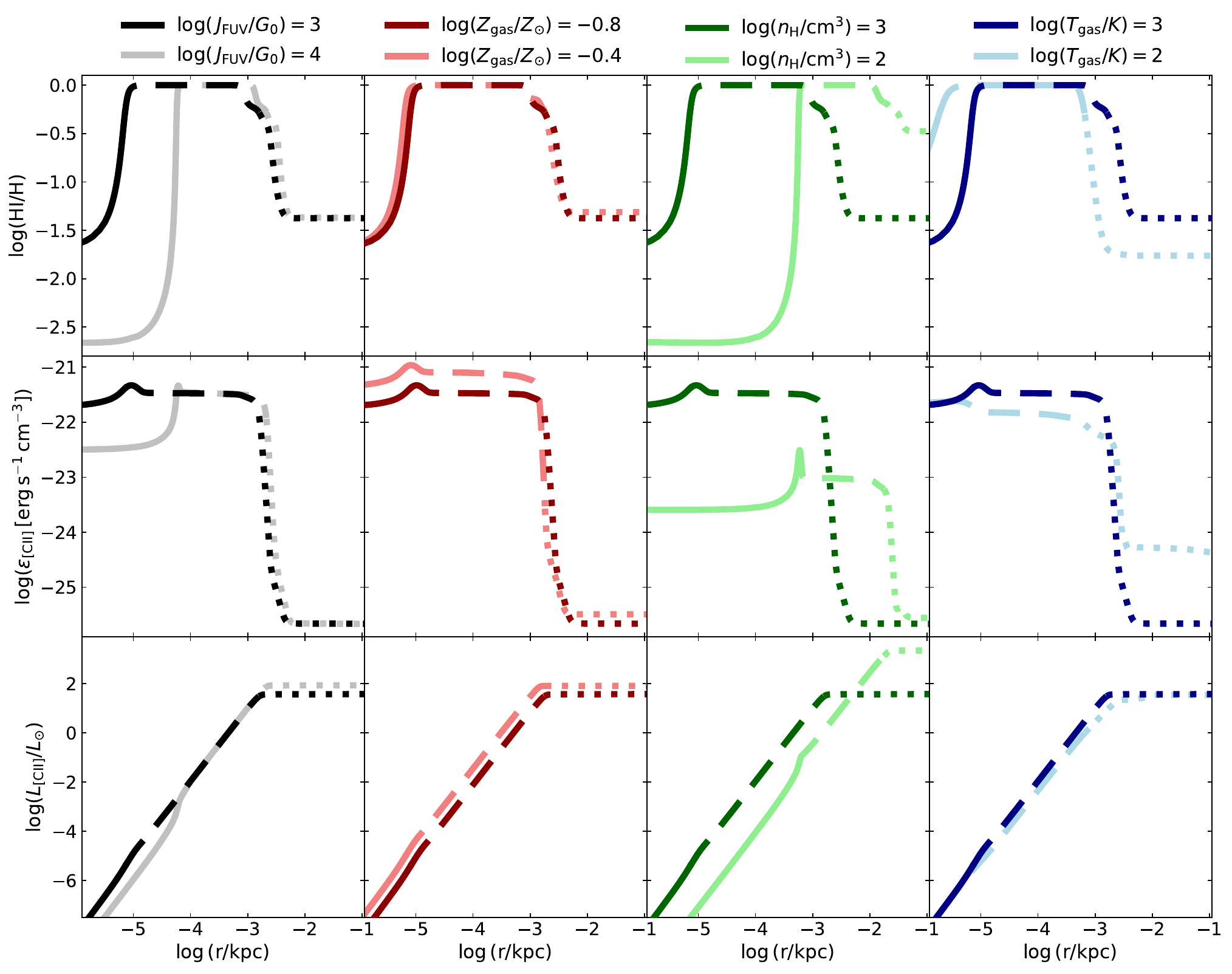}
\caption{Demonstration of how the \textsc{Cloudy} modeling of the [\ion{C}{2}] emission varies with the change of ISM parameters. From left to right column, we show the \textsc{Cloudy} modeling results by generating two variations of a simple physical parameter, $J_{\rm FUV}$, $Z_{\rm gas}$, $n_{\rm H}$, and $T_{\rm gas}$, respectively, where the baseline parameters are shown in dark-colored curves, and the parameter variations are shown in light-colored curves. From top to bottom row, we show key \textsc{Cloudy} output results, the abundance profile of atomic hydrogen HI/H, $\epsilon_{\rm [CII]}$, and cumulative $L_{\rm [CII]}$ vs. the depth into a gas cloud. Each continuous curve is characterized by three distinct zones in the \textsc{Cloudy} modeling based on the ionization state of hydrogen gas, which are the HII region (solid curve), HI-dominated region (dashed curve), and ${\rm H_{2}}$-dominated region (dotted curve).
\label{fig:Cloudy_12}}
\end{figure*}

We run \textsc{Cloudy} simulations over a grid of parameters (see Section~\ref{subsec:Cloudy}) to show how [\ion{C}{2}] emission depends on variations of relevant physical parameters, including $J_{\rm FUV}$, $Z_{\rm gas}$, $n_{\rm H}$, and $T_{\rm gas}$. Figure~\ref{fig:Cloudy_12} illustrates these dependencies, which show how the \textsc{Cloudy} output varies when a single parameter deviates from its fiducial value. The fiducial value of each parameter is determined based on \cite{Lagache18} and \cite{Liang24}, where $\log (J_{\rm FUV}/G_{0}) = 3$, $\log (Z_{\rm gas}/Z_{\odot}) = -0.8$, $\log (n_{\rm H} \, {\rm cm^{-3}}) = 3$, and $\log (T_{\rm gas}/{\rm K}) = 3$.

From Figure~\ref{fig:Cloudy_12}, we find that the HI-dominated region contributes to the majority of the [\ion{C}{2}] emission. Concerning the range of variations, the simulation results of $\epsilon_{\rm [CII]}$ and $L_{\rm [CII]}$ appear to be more sensitive to $n_{\rm H}$ than other parameters.

\section{Surface brightness images [\ion{C}{2}] from redshift 6 to 5}
\label{appendix:more_snapshots}

\begin{figure}[htp]
\centering
\includegraphics[width=0.99\textwidth]{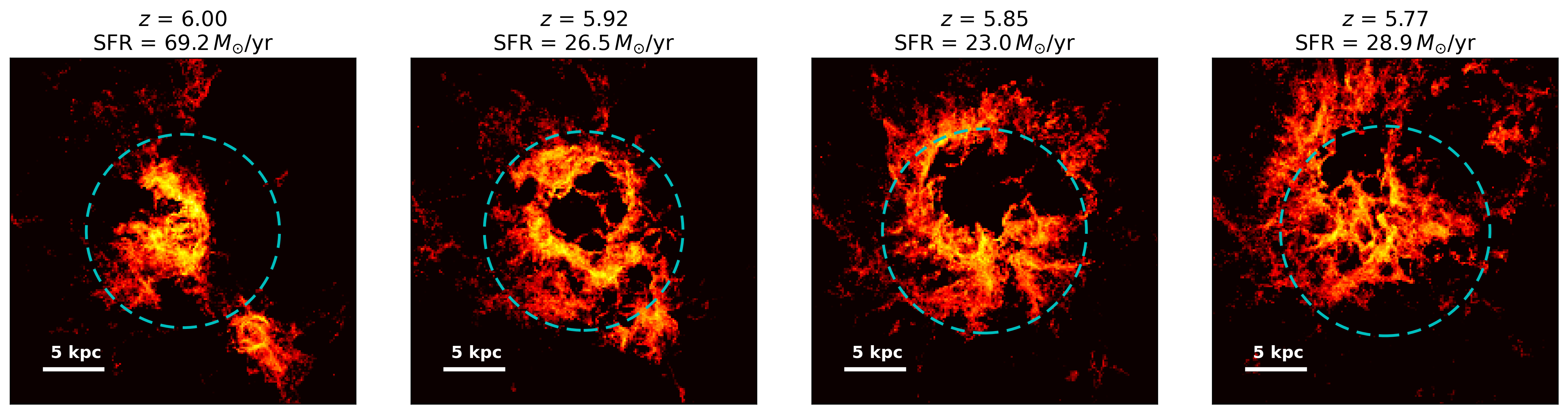}
\includegraphics[width=0.99\textwidth]{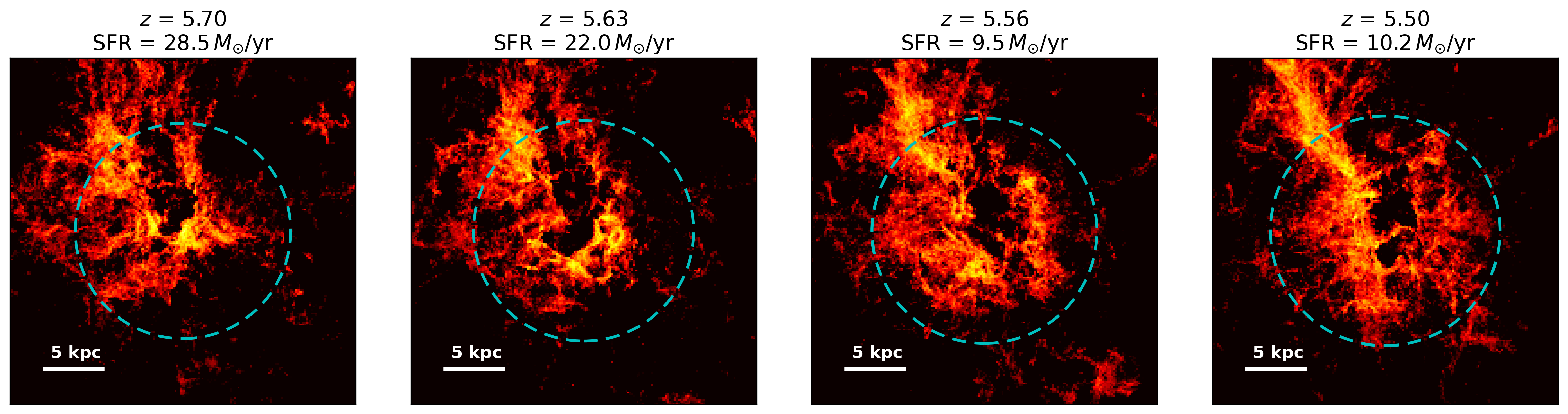}
\includegraphics[width=0.99\textwidth]{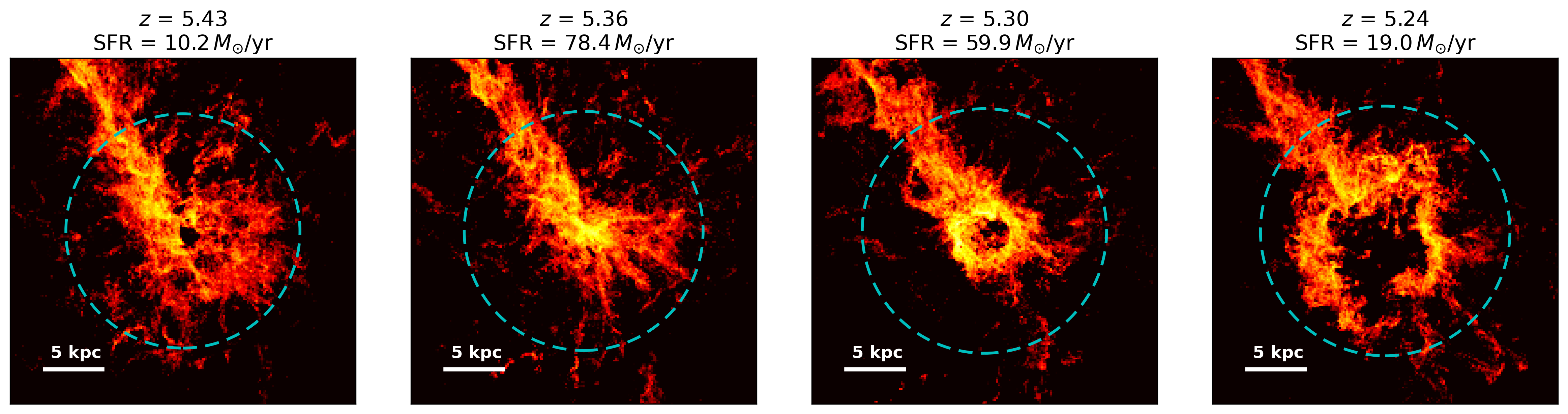}
\includegraphics[width=0.99\textwidth]{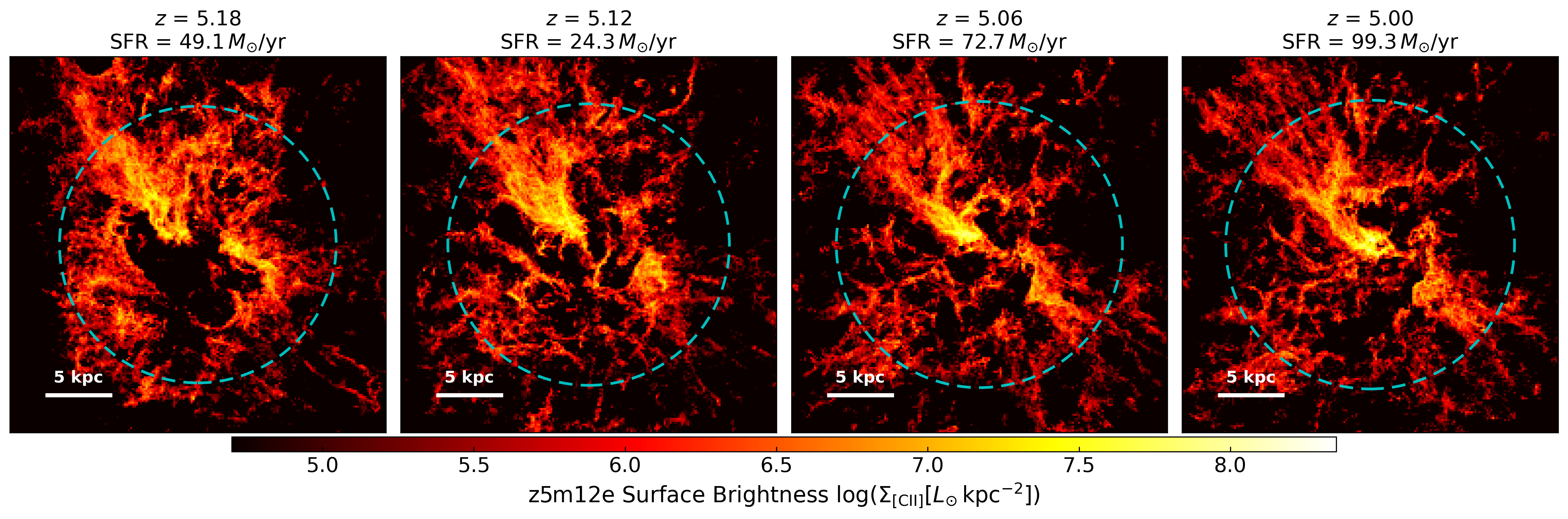}
\caption{All 16 snapshots of {\tt z5m12e} from $z = 6$ to 5. Each panel shows $\Sigma_{\rm [CII]}$ in the same format as Figure~\ref{fig:example_cii_halo}.}
\label{fig:combined}
\end{figure}

\begin{figure}[htp]
\centering
\includegraphics[width=0.99\textwidth]{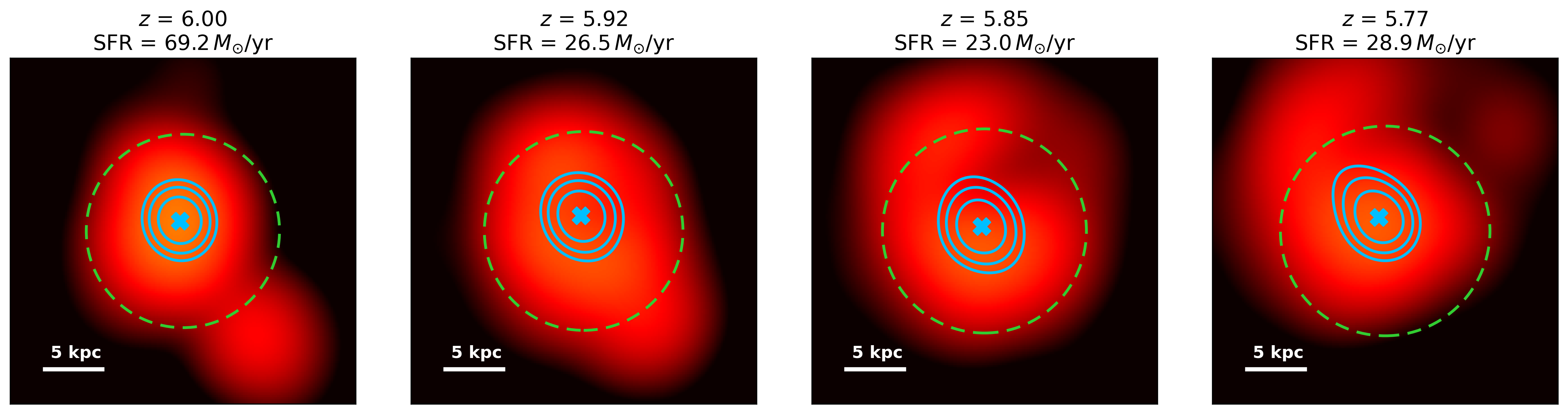}
\includegraphics[width=0.99\textwidth]{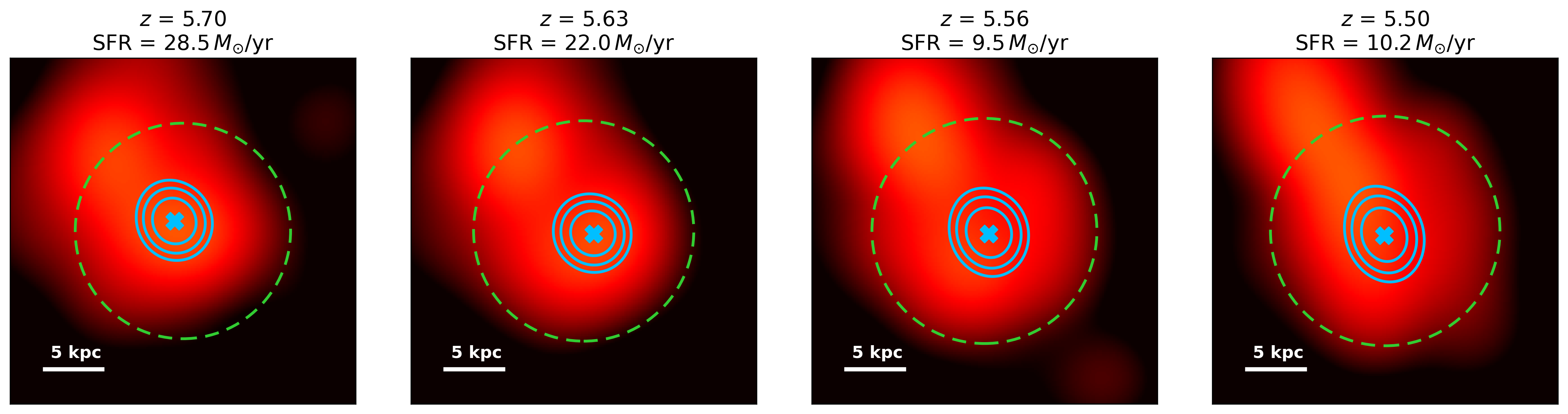}
\includegraphics[width=0.99\textwidth]{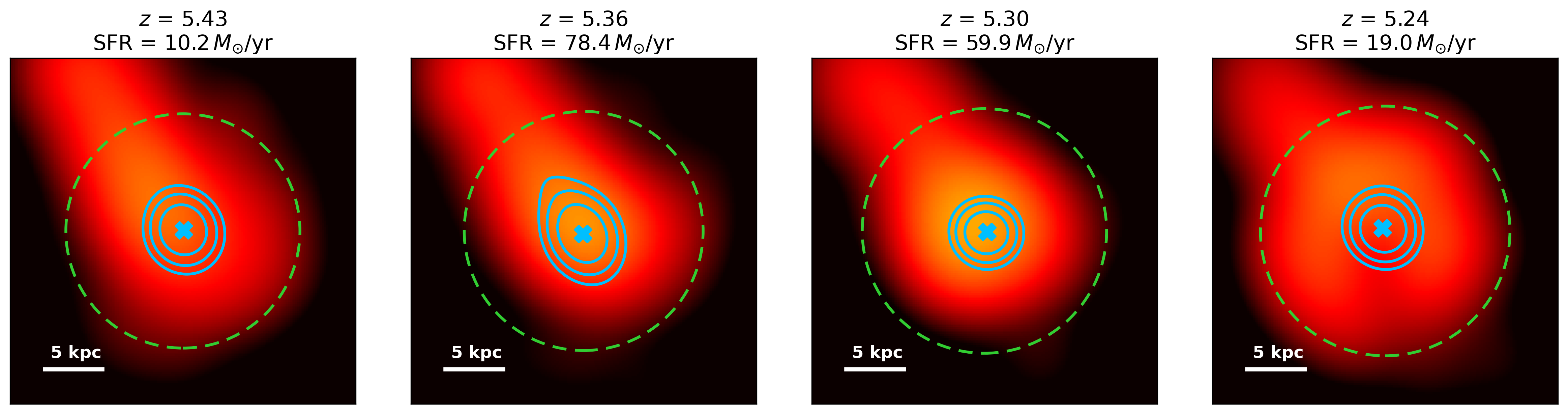}
\includegraphics[width=0.99\textwidth]{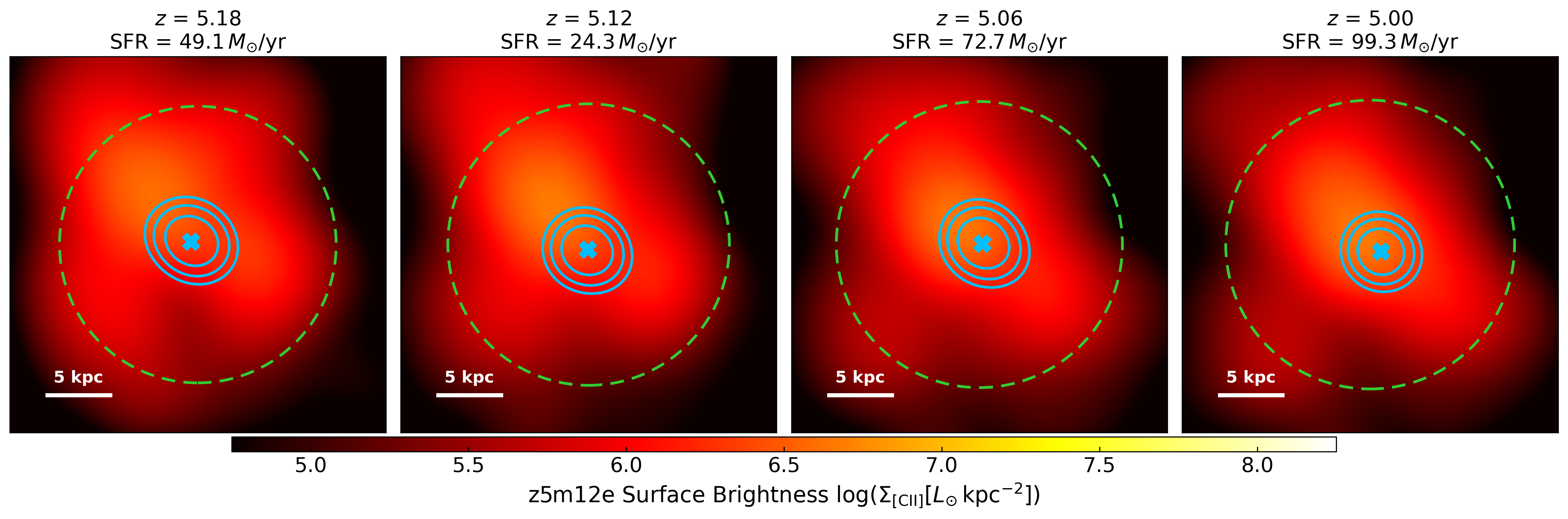}
\caption{{Similar to Figure~\ref{fig:combined}, but after convolving with a 2D Gaussian kernel comparable to the ALPINE beam size \citep{Fujimoto19,Fujimoto20}. In each panel, the green dashed circle marks the reference scale for our [\ion{C}{2}] halo analysis ($0.3 R_{\rm vir}$ or $\sim 10 \,$kpc), whereas the blue contours overplotted show the smoothed profile of UV continuum (the same beam size) out to half of its peak value, with the peak position indicated by the cross at the center.}}
\label{fig:smooth_combined}
\end{figure}

We show all 16 snapshots of {\tt z5m12e} from $z = 6$ to 5 in Figure~\ref{fig:combined} as an extended version of Figure~\ref{fig:example_cii_halo}. To facilitate direct comparison with observations, we further present Figure~\ref{fig:smooth_combined} where each raw [\ion{C}{2}] map is smoothed with a 2D Gaussian kernel ($\mathrm{FWHM} = 1 \arcsec$; $\sim 6\,$kpc at $z = 5$) matched to the ALPINE beam size \citep{Fujimoto19,Fujimoto20}, and we overlay the convolved UV continuum (the same beam size) profile to highlight the relative spatial extent of the [\ion{C}{2}] emission, which can build further intuition for observations. We note that at $z=6$ there is a satellite galaxy near the central galaxy (at the lower right of the first panel and outside $0.3R_{\rm vir}$), which has later merged into the central galaxy.

Comparing the intrinsic, unsmoothed [\ion{C}{2}] morphologies (Figures~\ref{fig:example_cii_halo} and \ref{fig:combined}) with the beam-smoothed ones in Figure~\ref{fig:smooth_combined} illustrates the suppression of [\ion{C}{2}] small-scale features, such as the clearance of [\ion{C}{2}]-emitting gas from the central region (e.g., at $z \approx 5.2$), due to the finite spatial resolution. As a result, our beam-smoothed [\ion{C}{2}] images become similar to ALMA observations. Comparing the [\ion{C}{2}] and UV morphologies smoothed with the same beam size illustrates that in most cases [\ion{C}{2}] is more extended than UV.

We note that in some snapshots an offset between the UV continuum and [\ion{C}{2}] peaks is discernible (e.g., at $z=5.12$) after beam smoothing. However, this offset remains relatively small compared to the spatial extent of [\ion{C}{2}] emission. Since the main purpose of this paper is to characterize the width and temporal evolution of 1D [\ion{C}{2}] profiles, this small offset would not significantly affect our results. Indeed, observational studies such as \cite{Fujimoto19} have found that stacking on the [\ion{C}{2}] or UV peak has little impact on the extended nature of [\ion{C}{2}] emission, and that the extended structure of [\ion{C}{2}] does not correlate with the [\ion{C}{2}]--UV offset. Therefore, we adopt the UV continuum peak as the common center for extracting the [\ion{C}{2}] and UV radial profiles.

\bibliography{sample701}{}
\bibliographystyle{aasjournalv7}

\end{document}